\documentstyle[12pt,aasms4]{article}

\newcommand{\msun}{$M/M_{\odot}\,$}

\begin{document}
 
\title{Theoretical Models for Classical Cepheids: II.  
Period-Luminosity, Period-Color and Period-Luminosity-Color Relations.}
 
\author{Giuseppe Bono\altaffilmark{1}, Filippina Caputo\altaffilmark{2},
Vittorio Castellani\altaffilmark{3}, and Marcella Marconi\altaffilmark{2}} 

\lefthead{Bono et al.}
\righthead{Theoretical Models for Classical Cepheids: III.} 

\altaffiltext{1}{Osservatorio Astronomico di Trieste, Via G.B. Tiepolo 11,
34131 Trieste, Italy; bono@oat.ts.astro.it}
\altaffiltext{2}{Osservatorio Astronomico di Capodimonte, Via Moiariello 16,
80131 Napoli, Italy; caputo@astrna.na.astro.it, marcella@cerere.na.astro.it}
\altaffiltext{3}{Dipartimento di Fisica, Univ. di Pisa, Piazza Torricelli 2,
56100 Pisa, Italy; vittorio@astr1pi.difi.unipi.it}

\begin{abstract}
 
We present and discuss theoretical predictions 
concerning the pulsational properties of Classical Cepheids. 
Masses and luminosities provided by stellar 
evolutionary calculations are used as input parameters of  
nonlinear, nonlocal and time--dependent 
convective pulsating 
models and accurate determinations of 
both the blue and red edge of the instability strip are derived,
together with theoretical light curves for a suitable grid of
models. The computations have been performed   
for  three different chemical compositions   
($Y$=0.25, $Z$=0.004; $Y$=0.25, $Z$=0.008; $Y$=0.28, $Z$=0.02),
taken as representative of Cepheids in 
the Magellanic Clouds (MCs) and in the Galaxy. 

Bolometric light curves have been
transformed into visual and near-infrared magnitudes and the 
intensity-weighted mean magnitudes of the pulsator 
over a full pulsation cycle 
($<M_V>$ and $<M_K>$, respectively) are obtained. We derive that 
either in the log$P$--$<M_V>$ and in the log$P$--$<M_K>$
planes the 
predicted edges of  instability strip  
are in excellent agreement with the 
observed distribution of Galactic and Magellanic Cepheids, providing 
a preliminary estimate of the distance to these galaxies. 
Moreover, we   
show that the models are in agreement with several empirical   
Period-Luminosity (PL) relations given in the literature, even though 
the theoretical distribution in log$P$--$<M_V>$ plane 
is better represented 
by a quadratic PL relation. We also show that both the zero-point and 
the slope of the predicted PL relations are significantly dependent 
on metallicity, with the amplitude 
of the metallicity effect decreasing at the 
longer wavelength.  At variance with several empirical 
suggestions appeared in the literature, we find that at fixed 
period the metal--rich pulsators should be {\it fainter} 
than the metal--poor ones. 

Tight Period-Luminosity-Color (PLC) 
relations are derived   
for both visual and near-infrared photometric bands. 
Also in this case the effect of  metallicity 
decreases with increased wavelength. 
>From a preliminary use of our relations to Magellanic Cepheids, 
we confirm that, within the statistical errors,  
the distance modulus obtained from different PL and PLC relations  
is marginally  correlated with 
the adopted relation, but  
the associated uncertainty decreases when infrared magnitudes 
are taken into account. 

Finally the whole pulsational scenario is briefly discussed in  
light of the adopted evolutionary framework. 
\end{abstract}
 
\noindent
{\em Subject headings:} Cepheids --- Galaxy: stellar content ---
Magellanic Clouds  --- stars: distances --- stars: oscillations

\pagebreak
\section{INTRODUCTION}
 
\noindent
The Cepheid Period-Luminosity (PL) relation has long played a central 
role in constraining Galactic and extragalactic distance 
scales. On very general grounds, the pulsation period 
is expected to be  tightly correlated with the luminosity, mass and
effective temperature of the pulsator. Furthermore, stellar evolution 
theory predicts, for each given original chemical composition,  
a relation between mass and luminosity. As a consequence, 
 at fixed chemical composition 
the Cepheid luminosity is expected to be   a function of period and 
effective temperature only. 
The outcome of this argument into the observative plane is a 
relation which sets the mean\footnote{The {\it mean} magnitude over a 
pulsation cycle may be either intensity-weighted $<M_j>$, or 
magnitude-weighted $(M_j)$. The discrepancy between these 
two different averages depend on the shape of the 
light curve. As a matter of example, we find that the difference 
$(M_V)-<M_V>$ is 0.016 mag at log$P$=1.0 and 0.070 mag at log$P$=1.7, 
whereas the difference $(M_K)-<M_K>$ ranges from 0.002 mag to 0.010 mag.} 
magnitude $<M_j>$ in a given    
bandpass as a function of period ($P$) and color index ($CI$), as 
given by 

$<M_j>$ = $\alpha$ + $\beta$ log$P$ + $\gamma$ [$CI$]

Dating back to the pioneering investigation by Sandage (1958), 
the existence and the estimate of the color term coefficient $\gamma$ 
in the 
period--luminosity--color [PLC] relation has been  
the subject of several thorough theoretical and observational 
analyses (see, e.g., Feast 1984, 1991, 1995; Caldwell \&  Coulson 1986; 
Laney \& Stobie 1986; Stothers 1988;  Caldwell \& Laney 1991; 
Sasselov et al. 1997). 
However, in the recent literature the effect of the color term has 
been widely neglected by adopting as observables only magnitudes 
and periods. In this framework, the distance modulus $DM$ 
is generally derived by adopting a period--luminosity [PL] 
relation which is often given in the linear form 

$\overline{<M_j>}$ = a + blog$P$, 

\noindent 
where $\overline{<M_j>}$ is the average of the mean magnitudes of 
Cepheids at a given period. 

Absolute magnitudes, and in turn distances, are usually derived according 
to Iben \& Renzini (1984) suggestion that 
the PL relation presents
a negligible dependence on the chemical composition. This 
assumption is adopted by the {\em Hubble Space Telescope} Key 
Project devoted to the identification of Cepheids in very distant 
galaxies (see the review by Freedman et al. 1997) and by the 
group working on the Cepheid 
calibration of SNe Ia (see, e.g., Saha et al. 1997). In the meantime, an 
unprecedented observational effort has been undertaken from ground-based 
and space telescopes for properly estimating both the zero-point and 
the slope of such a relation (Pierce et al. 1994). 

However, while a substantial improvement toward a sound calibration of the 
PL zero-point has been recently made by adopting the {\em HIPPARCOS} 
trigonometric parallaxes of Galactic Cepheids 
(Feast \& Catchpole 1997, hereafter FC97; Oudmaijer, 
Groenewegen \& Schrijver 1998), several recent empirical 
investigations suggest that the 
Cepheid PL relation presents a $non$ $negligible$ dependence 
on the metallicity. 
To quote some of the most recent results (see also 
Kochanek 1997 for a detailed review), 
Freedman \& Madore (1990) examined Cepheids in three fields of M31 
and found that the inferred 
distance moduli should be corrected according to 
a metallicity effect $\psi=DM_{true}-DM_{PL}$ of 0.32$\pm$0.21 mag/dex, 
whereas Gould (1994) from the same data derived 
$\psi$=0.88$\pm$0.16 mag/dex. 
More recently, Beaulieu et al. (1997) and
Sasselov et al. (1997) discussed the large photometric database of 
Magellanic Cepheids collected by the EROS project and 
found a factor 
$\psi$=0.44$^{+0.1}_{-0.2}$ mag/dex. 
Finally, Kennicutt et al. (1998) obtained
$\psi \sim$ 0.24$\pm$0.16 mag/dex from a 
differential analysis of Cepheids in
two fields of M101, a spiral galaxy in the Virgo cluster.  

Even though the previous empirical estimates of the metallicity 
correction seem to depend on the Cepheid sample and on the method 
adopted for disentangling reddening from metallicity effects, they 
agree in suggesting that the true distance of metal-rich Cepheids 
is {\it longer} than inferred from the PL relation. However, it is 
worth noticing that not all empirical relations support such a metallicity 
effect. As a fact, Caldwell \& Coulson (1986, hereafter CC86), 
by adopting optical photometric
data of Magellanic Cepheids, conclude that a decrease
in the metal content
implies a decrease in the distance moduli as based on the PLC relation
and an increase in the distance moduli based on the PL relation. 
Unfortunately, no firm indication can be derived from available 
theoretical predictions.  
Stothers (1988) found that a decrease in
the metal content implies a decrease in the distance moduli based on both
PLC and PL relations.     
Chiosi, Wood \& Capitanio (1993) predicted that the dependence of 
the absolute magnitude 
on the total metal content $\delta M_j$/$\delta Z$ ranges from  
$\sim$ -22.2 to $\sim$ 1.7 (depending on the adopted photometric 
band), and more recently 
Saio \& Gautschy (1997) derive that the PL relation should be  only marginally 
affected by metallicity. 

The metallicity factor is of fundamental importance as 
the PL uncertainty propagates 
into the extragalactic distance scale and into the evaluation of 
the Hubble constant $H_0$. Thus, one of the main aims 
in our investigation is to address the problem by using 
the extensive grid of Cepheid nonlinear models recently presented by 
Bono, Caputo \& Marconi (1998) and 
Bono, Marconi \& Stellingwerf (1998 hereinafter Paper I). The basic 
assumptions on the input physics and computing procedures, 
together with the adopted mass-luminosity relation, have been 
already described in Paper I. Note 
that the nonlinear approach presents manifold advantages in comparison 
with the linear one.  
First, a nonlinear regime which accounts 
for the coupling between pulsation and convection is needed 
to get theoretical predictions about both blue and red 
edges of the instability strip. On the contrary, in a 
linear-radiative-nonadiabatic  
approach only the blue edge can be predicted, whereas the red edge 
is either located according to a fixed temperature shift 
relative to the blue edge or by assuming an 
{\it ad hoc} efficiency of convective transport.     
This is a key point 
since the temperature width of the strip plays a relevant role for 
properly assessing the  properties of the PLC and PL relations. 
Moreover, full amplitude, nonlinear models supply useful 
predictions on the 
pulsational amplitudes to be compared with observations. On this basis 
it has been recently shown (Bono \& Marconi 1997) that updated 
evolutionary predictions, when related to pulsation, 
indicate a mild if not negligible efficiency of the core 
overshooting. 
Last, but not least, theoretical light curves  
can be directly compared with observations for 
constraining the physical parameters of variable stars 
(Wood, Arnold \& Sebo 1997) and/or for testing the 
accuracy and the consistency of the physical assumptions. 

Along this investigation, we start presenting in Section 2 
a comparison of present predictions
with published period-magnitude and period-color 
(PC)  
relations. In the 
same section the predicted edges of  
the instability strip in the period-magnitude plane 
are compared with the observed distribution of Galactic and 
Magellanic Cepheids. The analytical predicted PL, PC, and PLC relations 
are given in Section 3, while 
the comparison with Cepheids in the 
Galaxy and in the MCs is presented in Section 4. The last section deals 
with the summary of the main results. 


\section{Testing the theoretical scenario}

\subsection{Magnitudes}

\noindent
The adopted set of non linear, nonlocal and time-dependent convective models 
relies on four values of  the stellar mass (\msun=5.0, 7.0, 9.0, 11.0) and 
three chemical compositions 
($Y$=0.25, $Z$=0.004; $Y$=0.25, $Z$=0.008; $Y$=0.28, $Z$=0.02).
This set of  parameters was chosen in order to properly cover    
Cepheids observed both in the MCs and in  
the Galaxy. Helium abundances were fixed following  
$\Delta Y$/$\Delta Z$$\sim$2.5 and a primordial helium content 
$Y_P$=0.23. However, given the small (and uncertain) difference 
in metallicity between LMC and SMC, we adopted $Y$=0.25 for both 
galaxies. Finally, according to the discussion given in 
Bono \& Marconi (1997) and in Paper I,  
the mass-luminosity relation 
derived from "canonical" evolutionary models, i.e. 
with vanishing efficiency 
of convective core overshooting, has been adopted.

The results of the calculations are summarized in 
Fig. 1 where, for each assumed mass and luminosity, we plot 
the pulsation period as a function of the effective temperature  
throughout the instability strip.
Models which attain a stable nonlinear limit cycle in the fundamental 
(F) mode and/or in the first overtone (FO) are shown with filled and open 
symbols, respectively, while different chemical 
compositions are characterized by different symbols. We refer the
reader to Paper I 
for  the comparison between linear and nonlinear periods, 
as well as the comparison between our periods and other published
predictions. 

We find that the dependence of the period on the effective temperature
appears quite regular, with $\Delta$log$P$ $\sim$ 3.24$\Delta$log$T_e$
over the whole range of explored masses  and for
both F and FO pulsators. Moreover, the figure clearly shows that 
an increase in the metal content\footnote{In the following we will refer 
to "metallicity" even though for the $Z$=0.02 models this means 
also an increase of the helium content.} 
moves the instability strip toward 
cooler effective temperatures. As a consequence, for a given luminosity, 
an increase in metallicity causes an increase in the mean period of 
pulsators. Therefore, at fixed period, metal-rich pulsators are 
expected "on average" fainter 
than metal-poor ones. 

This metallicity effect is more 
evident in Fig. 2 which shows the mean bolometric magnitude 
$<M_{bol}>$ of fundamental pulsators as a function 
of the period. The solid curves plotted in the three panels
over the 
models are the quadratic least squares fits at the labelled
 metallicity. The data 
show that the 
magnitudes of the solar models present a 
rather 
linear dependence on log$P$, whereas the models at $Z$=0.008 and $Z$=0.004 
deviate more and more from linearity. This is evident from comparison
with the solar relation, which is plotted as a dashed line in the
upper two panels.
Moreover, data 
in Fig. 2 confirm that, at constant mass and luminosity, 
metal-rich pulsators have longer periods than metal-poor ones, thus 
indicating that at fixed period  
the bolometric magnitude is expected to increase with increased metallicity. 

In order to perform the comparison with observed photometric data, 
bolometric magnitudes have been 
transformed into the observative plane by adopting the grid of static 
atmosphere models provided by Castelli, Gratton \& Kurucz (1997a,b). On this 
basis we derive the theoretical predictions 
presented in Fig. 3 where the mean visual magnitude $<M_{V}>$ of 
F and FO pulsators is plotted against period. In this figure 
one may first note that 
FO pulsators with 5$M_{\odot}$ and 7$M_{\odot}$ run 
parallel to the sequence of F pulsators. This result 
gives a satisfactory explanation of the first overtone pulsators 
actually observed by MACHO and EROS collaborations (Alcock et al. 1995; 
Beaulieu et al. 1997; Sasselov et al. 1997). The predicted 
splitting between the two different modes is mainly due to the 
topology of the Cepheid instability strip which, 
unlike RR Lyrae variables, present a very narrow region in 
which the pulsators attain a stable limit cycle in both fundamental 
and first overtone modes. 

Furthermore, one should notice the evidence of an 
upper luminosity limit to the appearance of FO pulsators.  
This level is fixed by the luminosity $L^{IP}$ of the 
intersection point 
between the blue and red edge for first--overtone pulsation since,   
above this luminosity,  
the instability strip is populated  by F pulsators only.  
The evidence that 
FO pulsators are predicted with masses smaller 
than 7$M_{\odot}$ with $Z$=0.02 and smaller than 9$M_{\odot}$ 
at the lower metallicities shows  
that $L^{IP}$  decreases 
as the metal content increases. 

Figure 3 shows also 
some current empirical PL$_V$ 
relations for fundamental and first overtone Cepheids. 
The solid line refers to the relation recently 
derived by FC97 by adopting the slope for LMC Cepheids 
(Caldwell \& Laney 1991, hereafter CL91) 
and the zero-point from $HIPPARCOS$ data, while   
the dotted line is the 
relation derived by Madore \& 
Freedman (1991, hereinafter MF91) for LMC 
Cepheids. The dashed line is the relation obtained 
by the EROS collaboration (Sasselov et al. 1997) for a large 
sample of Magellanic Cepheids, but with 
the zero-point fixed according to FC97. 
Note that all the
above relations are based on averaged intensities transformed into
magnitudes, while the MF91 relation adopts averaged magnitudes. 
 
Finally, the dashed line plotted over FO pulsators is based on 
the empirical slope obtained by EROS for MC Cepheids, but with the 
FC97 zero-point shifted according to our predicted difference 
between fundamental and first overtone PL  
at $Z$=0.02. Even though more detailed observations 
are needed to assess the zero-point of these variables and, at 
the same time, pulsating models with lower masses are necessary for 
extending the theoretical predictions towards shorter periods, 
the agreement between present results and the empirical slope 
seems to support the suggestion by Sasselov et al. (1997) that 
first overtone Cepheids are safe distance indicators. The advantage 
deals with the narrow temperature range where they are 
pulsationally stable. On the other hand, they 
are characterized by small pulsation amplitudes and their identification 
requires a very good coverage of the light curve.   

A glance at Fig. 3 clearly suggests 
that for periods shorter than log$P\le$1.5 
the linear empirical PL$_V$ relations appear in reasonable 
agreement with the 
theoretical models at $Z$=0.004 and $Z$=0.008, i.e. for the 
typical average 
metallicities of MC Cepheids. 
Towards longer periods the empirical relations for fundamental pulsators  
tend to match the models close to the blue edge of the predicted 
instability strip.
 
Beyond such a comparison, we note that at each given luminosity 
the pulsation instability occurs 
over a finite range of effective temperatures. Therefore 
theoretical models supply  
firm constraints on the predicted range 
of periods.. This feature can be soundly compared  
with observational data, 
without the sacrifice of using $mean$ quantities as implicitly assumed 
in the PL relation. To this purpose the bottom panel of 
Fig. 4 shows the comparison in the log$P$--$<M_V>$ plane between  
the predicted edges of the instability strip for fundamental pulsators with $Z$=0.008
(solid line),  and the reddening--corrected 
$<V>_0$ magnitudes and periods of 47 SMC Cepheids collected by  
Laney \& Stobie (1994, hereinafter 
LS94). The agreement between observations 
and the predicted edges of the instability strip 
is quite impressive, 
provided that a distance modulus $DM_{SMC}$=19.1, with an 
uncertainty of $\pm$0.1,  is adopted. 

The top panel of Fig. 4 shows the comparison between 
theoretical predictions with $Z$=0.008 and 45 LMC Cepheids 
($<V>_0$ and periods from LS94). Once again the agreement 
between theory and observations is very good, 
if a distance modulus of  
$DM_{LMC}$=18.6 ($\pm$0.1) is adopted. 
However, it is noteworthy that the period of 
the brightest  
LMC variables 
tend to be somehow shorter than the predicted blue edge. 
This effect could be a failing of the theoretical scenario, but we suggest 
that such a minor disagreement might be due to the occurrence of luminous 
pulsators with metallicity lower than the commonly assumed value 
of $Z$=0.008. In fact, 
these bright Cepheids appear in good agreement with the blue 
edge of the $Z$=0.004 predicted 
strip (dashed lines), a suggestion which is 
supported from recent spectroscopic observations of 9 LMC Cepheids
(Luck et al. 1998) showing that the metallicity ranges from
$Z\sim$ 0.006 to $Z\sim$ 0.013. Moreover, it should be 
noticed that the occurrence of metal-poor Cepheids 
is favored in comparison with metal-rich ones (for 
the former the time spent 
inside the instability strip is longer), as well 
as that the pulsation amplitudes of the models with 11$M_{\odot}$ are 
steadly decreasing going from the blue to the red edge of the instability 
strip. Thus, the probability to detect long-period 
Cepheids is higher close to blue edge than to the red edge. Finally, it 
should be mentioned the thorny problem of undetected binary companions 
which could seriously affect the mean magnitude of long-period Cepheids 
(Szabados 1997). 

Figure 5 shows the comparison between models and predicted edges at 
$Z$=0.02 (solid line) with two different samples of Galactic Cepheids. 
The bottom panel show LS94 data, while the top panel refers to 
data by Gieren, Fouqu\`e \& Gomez (1997, hereinafter GFG97). One may 
note that 
the occurrence of luminous variables with metallicity significantly 
lower than the 
currently adopted mean value (dashed lines refer to $Z$=0.008) 
seems present also in our Galaxy. Indeed, recent 
spectroscopic observations by Fry \& Carney (1997) for a 
sample of 25 Galactic Cepheids show metallicities in the range of
$Z$=0.008 to $Z$=0.024. However, 
the longest period Cepheids in both samples are SV Vul ($\log{P}=1.65$),
GY Sge ($\log{P}=1.71$) and S Vul ($\log{P}=1.84$), and only for SV Vul 
we have some estimate of metallicity ([Fe/H]$\simeq{0.06}$
by Fry \& Carney 1997; [Fe/H]$\simeq{0.23}$ 
by Luck \& Lambert 1985). These metallicities
are actually not supporting our ``sub-solar'' hypothesis, rather
confirming  that the radius determinations overestimate the true radii
of these stars (see GFG97). On the other hand, the absolute magnitudes listed
by LS94, which are based on ZAMS fitting, appear much more in agreement 
with the theoretical predictions, also in view of the adopted 
Pleiades distance modulus (5.57 mag), which 
is a bit larger than the value suggested
by Hipparcos data (5.33 mag). 

The previous comparison between theory and observations 
can be repeated by adopting 
infrared magnitudes instead of visual magnitudes. 
Figure 6 shows that the 
predicted distribution in the log$P$-$<M_K>$ plane 
confirms early suggestions (McGonegal et al. 1983; LS94) 
that the use of infrared magnitudes is 
a powerful method for reducing and, to some extent, for 
avoiding the spurious effects on the PL relation introduced both by the 
intrinsic width of the instability strip 
and by its dependence 
on the metal content. 
As a matter of fact, following the dependence of the bolometric 
correction on the effective temperature, the pulsators characterized by 
different metal contents scale almost linearly with the logarithmic 
period, i.e. a decrease of the effective temperature (increase of the 
period) implies a marked decrease of the $K$ magnitude. 
Therefore, both metal-poor and metal-rich  
theoretical PL$_K$ relations show a much more linear behavior, with only 
slightly different slopes and zero-points. 
Data in Fig. 6 also shows that the 
theoretical models appear in excellent 
agreement with the empirical relations 
provided by MF91, LS94, and GFG97. No comparison can be 
performed for FO pulsators since we still lack empirical relations 
in the $K$ band.   

Concerning the comparison of the predicted edges of the instability strip    
with measured infrared magnitudes of Magellanic Cepheids, 
Fig. 7 shows again a remarkable agreement,  for  
the MC distance moduli already  adopted in Fig. 4 
($DM_{SMC}$=19.1$\pm$0.1 and $DM_{LMC}$=18.6$\pm$0.1). As for Galactic 
Cepheids, Fig. 8 confirms the agreement between theoretical predictions and 
observations, except for the three longest period variables already discussed 
in the comments on Fig. 5.   

\subsection{Colors}

\noindent
Observational PC relations are generally given as linear relations connecting 
the period and the average of the mean\footnote{The three methods of 
deriving the "mean" 
color over the pulsation cycle use either ($m_j-m_i$), the average over the 
color curve taken in magnitude units; or $<m_j-m_i>$, the 
mean intensity over the color curve transformed into magnitude;
or $<m_j>$-$<m_i>$, the difference of the mean intensities transformed into 
magnitude, performed separately over the two bands. Generally, 
($m_j-m_i$) colors are redder than $<m_j>$-$<m_i>$ 
colors, with the  difference
depending on the shape of the light curves. Indeed, we find that the 
difference  
($B-V$)-[$<B>-<V>$] ranges from 0.018 mag at log$P$=1.0 to 
0.08 mag at log$P$=1.7, whereas the difference ($V-K$)-[$<V>-<K>$] is 
in the range of 0.014 mag, to 0.060 mag.} colors of Cepheids at a fixed 
period
  
$\overline{<CI>}$ = $A$ + $B$log$P$

By adopting the atmospheric models provided by 
Castelli, Gratton \& Kurucz (1997a,b), we derived the $<B>-<V>$ colors 
of F and FO pulsators with the three different metallicities. Figure 9 
shows the predicted colors as a function of the period, in comparison 
with the empirical PC relations obtained by CC86 for Galactic and 
Magellanic Cepheids. The principal features of Fig. 9 are: {\it i}) 
in contrast with the period--luminosity diagram,
the distribution of FO pulsators cannot be clearly identified as 
they present almost the same colors of F pulsators; {\it ii}) 
within the observational uncertainties, the 
empirical PC relations for Magellanic 
Cepheids are in good agreement with  
the $Z$=0.004 and $Z$=0.008 models; and {\it iii}) 
the empirical PC relation for Galactic 
Cepheids is systematically bluer than the $Z$=0.02 models 
(filled circles). As shown in 
the bottom panel of  
Fig. 10, the same conclusions are derived if the empirical PC 
relations by LS94 are adopted. However, one should consider the 
theoretical uncertainty at the edges of the instability strip 
($\pm$ 150 K, including the contribution of 
microturbolence velocity and/or individual 
element distribution) and, more importantly, the strong dependence 
of theoretical colors on the adopted set of static atmosphere models. 
As a whole, we evaluate that the change of the color at the blue edge 
may grow up to  
$\pm$ 0.05 at $Z$=0.004 and $\pm$ 0.08 at $Z$=0.02, while at the 
red edge we obtain $\pm$ 0.08. 

Following the arguments already discussed for the period-magnitude 
plane, we present in Fig. 11 the comparison between LS94 data and the 
predicted edges of the instability strip at the different metallicities. 
Taking into account both theoretical and observational uncertainties, as 
well as the already suggested small dispersion 
in the metallicity of Galactic and LMC Cepheids, the agreement 
between theoretical predictions and observations is reasonably good. 

We can now perform the same analysis for the $<V>-<K>$ color. The top 
panel of Fig. 10 shows the comparison between theoretical models 
at the different metallicities and the empirical PC relations by LS94. 
Once again we find a reasonable agreement for Magellanic Cepheids, whereas 
our $Z$=0.02 models appear bluer than observed Galactic Cepheids. As 
for the comparison between LS94 data and theoretical edges (see 
Fig. 12), some interesting points are worth of notice: {\it i}) even though 
the $<V>-<K>$ color is less affected by line blanketing and metallicity, 
the observed dispersion is larger than in the log$P$-($<B>-<V>$) plane, and 
{\it ii}) given the strong dependence of infrared colors on 
the effective temperature, the theoretical uncertainty of 
colors at the blue edge 
is here $\pm$0.10 with $Z$=0.004 and $\pm$0.12 with $Z$=0.02. The red edge 
is even more affected, with  
$\pm$0.16 at $Z$=0.004 and $\pm$0.19 at $Z$=0.02. As a conclusion, the 
intrinsic scatter of observations and the uncertainties of theoretical colors 
seem to prevent a sound comparison, almost supporting the suggestion by 
LS94 that model colors in the optical bands are much more reliable 
than the infrared ones.


\section {Analytical PL, PC, and PLC relations}

\subsection{Period-Luminosity}
\noindent
According to the procedure largely adopted in the current literature,
in this section we discuss the theoretical results concerning  
the average value $<\overline{M_j}>$ predicted for Cepheids with  
a given period. Figure 13 
shows our theoretical  log$P$--$<\overline{M_V}>$  
relation, as obtained for each 
adopted metallicity by a quadratic least squares fit of theoretical 
pulsators. Over the whole range of fundamental periods we find:
$$Z=0.004 \ \ \ \overline{<M_V>}=-0.75(\pm 0.06)-4.21(\pm 0.05)\log{P}
		+0.66(\pm 0.02)({\log{P}})^2\eqno(1a)$$
$$Z=0.008 \ \ \ \overline{<M_V>}=-0.84(\pm 0.03)-3.97(\pm 0.02)\log{P}
		+0.63(\pm 0.02)({\log{P}})^2\eqno(1b)$$
$$Z=0.02 \ \ \  \overline{<M_V>}=-1.37(\pm 0.06)-2.84(\pm 0.06)\log{P}
		+0.33(\pm 0.02)({\log{P}})^2\eqno(1c)$$

\noindent
with a standard deviation  $\sigma_V\sim$ 0.27 mag. As far as FO models 
are concerned, we find that the PL$_{<\overline{M_V}>}$ relation 
is well approximated by a unique linear relation:  

$$\overline{<M_V>}=-2.01(\pm 0.14)-2.93(\pm 0.14)\log{P}\eqno(1d)$$
\noindent 
with $\sigma_V\sim$0.13 mag. 

Wishing to constrain the theoretical results into a linear approximation, 
we show in Fig. 14 the least squares solutions for fundamental models with 
masses 5$M_{\odot}$--9$M_{\odot}$ and 9$M_{\odot}$--11$M_{\odot}$. 
For periods shorter than log$P\sim$ 1.4 we obtain 

$$Z=0.004 \ \ \overline{<M_V>}=-1.20(\pm 0.03)-3.04(\pm 0.05)\log{P}\eqno(1e)$$
$$Z=0.008 \ \ \overline{<M_V>}=-1.32(\pm 0.04)-2.79(\pm 0.06)\log{P}\eqno(1f)$$
$$Z=0.02 \ \  \overline{<M_V>}=-1.62(\pm 0.03)-2.22(\pm 0.04)\log{P}\eqno(1g)$$
  
\noindent
with with $\sigma_V\sim$0.13 mag, while for periods longer than log$P\sim$ 1.4 
we get 

$$Z=0.004 \ \ \overline{<M_V>}=-2.42(\pm 0.02)-2.10(\pm 0.04)\log{P}\eqno(1h)$$
$$Z=0.008 \ \ \overline{<M_V>}=-2.47(\pm 0.04)-1.92(\pm 0.06)\log{P}\eqno(1i)$$
$$Z=0.02 \ \  \overline{<M_V>}=-2.60(\pm 0.04)-1.56(\pm 0.05)\log{P}\eqno(1l)$$

\noindent
with with $\sigma_V\sim$0.11 mag. As a general, either by quadratic or 
linear least-squares solutions, both the slope and the zero-point of 
theoretical PL$_{<\overline{M_V}>}$ relation are {\it dependent on 
the metallicity}. 

As for near-infrared magnitudes, 
we find that the theoretical results (see Fig. 15) 
can be approximated from linear relations. 
For fundamental pulsators we derive 

$$Z=0.004 \ \ \overline{<M_K>}=-2.66(\pm 0.08)-3.27(\pm 0.09)\log{P}\eqno(2a)$$
$$Z=0.008 \ \ \overline{<M_K>}=-2.68(\pm 0.08)-3.19(\pm 0.09)\log{P}\eqno(2b)$$
$$Z=0.02 \ \  \overline{<M_K>}=-2.73(\pm 0.07)-3.03(\pm 0.07)\log{P}\eqno(2c)$$
\noindent
with $\sigma_K$=0.12 mag, while for FO models we get  

$$\overline{<M_K>}=-3.12(\pm 0.05)-3.44(\pm 0.05)\log{P}\eqno(2d)$$
\noindent
with $\sigma_K$=0.05 mag.

As a whole, all the above analytical relations show 
that both the zero-point and the slope of the 
predicted PL relations are significantly  
dependent on metallicity. At the same time, they confirm 
that for log$P \ge$ 0.5 the metal-rich Cepheids are fainter 
than metal--poor ones, at a given period. 
The amplitude of this effect decreases in the short period range and 
with increased wavelength. Specifically, we obtain that the change in 
luminosity ranges from $\delta M_V\sim$ 0.6$\delta$log$Z$ and  
$\delta M_K\sim$ 0.2$\delta$log$Z$ at log$P\sim$1.0   
to $\delta M_V\sim$ 1.1$\delta$log$Z$ and 
$\delta M_K\sim$ 0.6$\delta$log$Z$ at log$P\sim$2.0 

\subsection{Period-Color}

The PC relations are often adopted for estimating the individual reddening of 
Cepheids (Pel 1985; Feast 1991) or the color shift of Cepheid samples 
characterized by different metallicities (LS94). Moreover, PC relations and 
color-temperature transformations are often coupled for evaluating the 
mean temperature of the variables (Pel 1980; Fry \& Carney 1998). 

The linear relations we derive for optical colors of 
fundamental pulsators are  
$$Z=0.004 \ \ \overline{<B>-<V>}=0.37(\pm 0.04)+0.30(\pm 0.05)\log{P}\eqno(3a)$$
$$Z=0.008 \ \ \overline{<B>-<V>}=0.36(\pm 0.05)+0.38(\pm 0.05)\log{P}\eqno(3b)$$
$$Z=0.02 \ \  \overline{<B>-<V>}=0.36(\pm 0.02)+0.52(\pm 0.02)\log{P}\eqno(2c)$$
\noindent
with $\sigma_{B-V}$=0.10 mag, while for FO models we get

$$\overline{<B>-<V>}=0.42(\pm 0.06)+0.19(\pm 0.07)\log{P}\eqno(3a)$$
\noindent
with  $\sigma_{B-V}$=0.07 mag. 

As for near-infrared colors, we derive for fundamental pulsators 

$$Z=0.004 \ \ \overline{<V>-<K>}=1.19(\pm 0.07)+0.55(\pm 0.09)\log{P}\eqno(4a)$$
$$Z=0.008 \ \ \overline{<V>-<K>}=1.10(\pm 0.08)+0.70(\pm 0.09)\log{P}\eqno(4b)$$
$$Z=0.02 \ \  \overline{<V>-<K>}=0.95(\pm 0.04)+1.00(\pm 0.04)\log{P}\eqno(4c)$$
\noindent
with $\sigma_{V-K}$=0.19 mag, while for FO models we get

$$\overline{<V>-<K>}=1.11(\pm 0.09)+0.51(\pm 0.09)\log{P}\eqno(4d)$$
\noindent
with  $\sigma_{V-K}$=0.09 mag. 

\subsection{Period-Luminosity-Color}
                                                    
Let us remind that all the   
PC and PL relations are "statistical" solution. Our predicted PL relations
have been derived under the hypothesis that
the instability
strip is uniformly populated; therefore, any particular Cepheid distribution
within the instability strip would obviously
affects the result. Moreover, the finite width of the instability 
strip causes, mainly in the $V$-band, 
an intrinsic scatter in any period--magnitude or period--color relation. 

Obviously, this does not happen if also the mean color of each 
pulsator is taken into account since, as stated at the beginning of 
this paper, only a PLC relation can soundly account for the 
Cepheid pulsational 
properties (see also Feast 1995). Notwithstanding such straightforward 
arguments, the existence and estimate of the color coefficient has 
been a thorny observational problem. Here we wish only to recall 
that earlier 
estimates of the $\gamma$ coefficient for the $<B>-<V>$, based 
on LMC Cepheids, ranged from 
5 (Brodie \& Madore 1980), to 2.7 (Martin, Warren \& Feast 1979), to 
almost zero (Clube \& Dawe 1986). Only ten years ago the question 
was definitively settled 
by Laney \& Stobie (1986) who demonstrated that the color term is a real 
property caused by the intrinsic width of the strip and not by 
differential reddening.  

In order to supply theoretical PLC relations, 
we performed the least square solutions through 
our models at fixed metallicity, 
by assuming as independent variables log$P$ and 
the color. Table 1 gives the 
derived values for  $\alpha$, $\beta$, and $\gamma$ coefficients 
for fundamental pulsators, together with  
the standard deviations $\sigma$. Owing to the small number 
of FO pulsators, the analytical PLC relations have been derived irrespectively 
of the different chemical compositions. We obtain
$$<M_V>=-2.73(\pm 0.08)-3.26(\pm 0.10)\log{P}+1.72(\pm 0.31)(<B>-<V>)\eqno(5a)$$
$$<M_K>=-3.70(\pm 0.01)-3.71(\pm 0.01)\log{P}+0.52(\pm 0.01)(<V>-<K>)\eqno(5b)$$
\noindent
with $\sigma_V$=0.08 mag and $\sigma_K$=0.01 mag. 

As shown in Fig. 16, the predicted PLC relation based on 
$<B>-<V>$  should provide a very precise evaluation 
(to within $\sim$ 0.05 mag) 
of the absolute magnitude $<M_{V}>$ of a 
Cepheid with known period and color, 
{\it once the dependence on metallicity is 
properly taken into account}. 
Note also that the predicted PLC relations 
suggest that 
metal-rich Cepheids are now {\it brighter} than metal-poor ones, 
at constant period and $<B>-<V>$ color, 
in agreement with earlier observational (CC86) and 
theoretical (Stothers 1988) suggestions. 

As for the comparison between the coefficients in Table 1 and recent 
empirical estimates, we find a good agreement for what concerns the 
value of $\beta$ (see CC86, CL91), whereas the comparison on 
the zero-point ($\alpha$) and color coefficient ($\gamma$) is hampered by the method adopted for 
treating observational errors. In fact, CL91 find two different solutions 
for 42 SMC Cepheids ($\gamma=$2.34$\pm$0.19 and 3.35$\pm$0.43, against 
the predicted value of 2.79$\pm$0.07 at $Z$=0.004) and for 27 LMC 
Cepheids ($\gamma=$2.19$\pm$0.23 and 2.75$\pm$0.29, 
against the predicted value 
of 2.83$\pm$0.06 at $Z$=0.008). Finally, for Galactic Cepheids, the 
available empirical estimates for $\gamma$ range from 2.52 (Sandage \& Tammann 1969) 
to 2.13 (Caldwell \& Coulson 1987), somehow lower than the predicted 
value 3.27$\pm$0.18 at $Z$=0.02.           
   
Moving to infrared colors, Fig. 17 shows that 
the predicted PLC relation based on
$<V>-<K>$ colors has a reversed metallicity effect, 
with the metal-rich 
variables fainter than the metal-poor ones at
fixed period and color. However, the amplitude of 
the metallicity effect is largely reduced, 
confirming once again that infrared magnitudes and colors should 
considerably help to establish accurate distance scales also in 
absence of precise metallicity estimates.     

\section{Galactic and Magellanic Cloud Cepheids}

A first comparison between present theoretical 
PL relations at solar chemical  
composition and observations is 
presented in Fig.s 18 and 19 
for the two samples analyzed by LS94 (bottom) and GFG97 (top). 
>From Fig. 18 it is clear that long--period Cepheids 
appear more and more 
brighter in comparison with the predicted relation with $Z$=0.02 
(eq. (1c), solid line). This evidence, which 
causes the difference in the slope between theoretical and empirical 
relations, could be understood if we admit that 
the long--period (brighter) 
Cepheids are metal-poor variables with $Z\sim$ 0.008 (dotted line). On the 
other hand, the uncertainty related to the absolute magnitude of Galactic 
Cepheids has been already discussed in Sec. 2.1. 
If infrared magnitudes are taken into account, see Fig. 19,  
then the narrower width of the strip and 
the milder dependence on 
metallicity yield a fair agreement between theoretical predictions 
(eq. (2c), solid line) and  
observed Cepheids (and empirical $PL_K$ relations). 

A further not marginal support to the reliability of the present theoretical 
scenario is presented in Fig. 20 where the predicted infrared PLC 
relation is compared with Galactic Cepheids. 
Since the PLC relation provides a $unique$ solution for 
the magnitude of a Cepheid with known period and color, 
and given the marginal dependence on metallicity 
of the infrared relation, 
the agreement between theory and observed data turns out to be almost 
perfect. 
   
The full application of our predictions to Cepheids in our Galaxy 
and  external galaxies is beyond the purpose of the present 
paper. However, it seems worth underlining some problems connected 
with the determination of the distance modulus based on 
fiducial PL and PLC relations. 
Table 2 summarizes the SMC and LMC true distance moduli, with the 
related uncertainty, as derived from  
LS94 data (either magnitudes and colors are
given by the authors corrected for reddening) by adopting 
our predicted PL and PLC relations with the labeled 
metallicity. Even though
all the results agree to 
within the errors, we note the evidence of significative trends. 
In particular, the mean values reported in Table 2 
 increase from optical to infrared  relations, while the 
standard deviations tend to decrease. 
We wish to recall that the derived distance moduli are
significantly affected by both the finite width
of the instability strip (which shrinks
when moving to infrared)
and  the metallicity effects (which decrease with
increasing wavelength).  

Let us suppose to be 
unaware of LMC metallicity and to derive 
the LMC distance modulus from the relations 
given in this paper for different chemical compositions. 
The resulting values are plotted 
in Fig. 21 (dots) together with, just for reference,  
the distance modulus given by the FC97 relation 
($DM_{LMC}$=18.70, dashed line)\footnote{Oudmajer et al. 1998 have recently 
revised this value suggesting 18.53$\pm$0.08. However, in the 
present context the FC97 estimate is taken as a reference value and the 
aftermaths of the metallicity effect on distance determinations are not 
affected.}. The figure shows quite clearly 
the dependence of the distance determination in different photometric 
bands on metallicity. In particular, the estimate from the 
infrared PLC relation is  
rather constant ($\delta DM$/$\delta$log$Z\sim$ -0.08),
whereas the value from the optical PLC relation increases for 
increasing metal content ($\delta DM$/$\delta$log$Z\sim$ +0.40). 
On the other hand, the PL relation yields a distance modulus which 
decreases at the larger metallicities  
($\delta DM$/$\delta$log$Z\sim$ -0.75 and -0.33 in the $V$ and 
in the $K$ band, respectively). 

\section{Summary}

Along this paper we have driven the attention on the evidence 
that convective pulsating models 
predict a significative dependence of the PL 
relation
on metals, mainly for optical magnitudes. 
We show that as the metallicity increases 
the instability strip moves toward redder colors, causing that 
metal-rich pulsators have on average 
longer periods than metal-poor ones 
with the same mass and luminosity.  
This yields that {\it at fixed period and for
uniformly populated instability strip} the mean bolometric magnitude 
increases at larger metallicities. 
 
Due to the dependence of the bolometric correction on the 
effective temperature, this effect is further enhanced when 
visual magnitudes are taken into account. 
At fixed period metal--rich Cepheids 
with log$P\ge$ 0.5 are expected  {\it fainter} 
than metal--poor ones. Owing to the non linear behavior of the predicted 
optical PL$_V$ relation, the metallicity effect increases in the 
long period range, with the difference in magnitude  
varying from     
$\delta M_V\sim$ 0.6$\delta$log$Z$ at log$P$=1.0 to 
$\delta M_V\sim$ 1.1$\delta$log$Z$ at log$P$=2.0.

At the longer wavelengths, our models confirm observational evidence 
concerning the linearity 
and the mild 
dependence on metallicity  
of the infrared PL$_K$--relation. Indeed, we derive  
$\delta M_K\sim$ 0.2$\delta$log$Z$, at log$P$=1.0 and 
$\delta M_K\sim$ 0.6$\delta$log$Z$, at log$P$=2.0.
 
As a whole, the predicted dependence 
on metallicity 
of both the zero--point and slope is a warning against the 
adoption of a {\it universal   
PL relation} for Galactic and extra--galactic Cepheids.   
Moreover, even though the metallicity is properly taken into 
account, the intrinsic scatter in the PL relations due 
to the finite temperature width of the instability strip 
prevents accurate distance determinations, mainly in the optical 
range. 
 
The uncertainty  in distance evaluations associated with the 
finite width of the instability strip, 
and with particular distributions of Cepheids within the 
instability strip, is strongly limited in distance estimates based on  
PLC relations. Our predictions 
suggest that the absolute magnitude 
of Cepheids with known metallicity, period, and color can be inferred 
with an accuracy of the order of  
$\pm$ 0.05 mag. 
Moreover, also in this case we find that the amplitude of 
the metallicity effect is greatly reduced by adopting infrared 
magnitudes and colors.

In order to make clear the dependence of distance determinations on 
the metallicity, we evaluate the LMC distance modulus by assuming to 
be unaware of its metallicity and by using the predicted 
PL and PLC relations at the different chemical abundances. We show that 
only the estimate based on  
the infrared PLC relation is
{\it bona fide} independent of the adopted metallicity 
($\delta DM$/$\delta$log$Z\sim$ -0.08), 
whereas from the optical PLC relation we get 
$\delta DM$/$\delta$log$Z\sim$ +0.40. 
On the other hand, the distance modulus based 
on PL relations decreases with increasing 
metallicity according to $\delta DM$/$\delta$log$Z\sim$ -0.75 and -0.33 
for the $V$ and
the $K$ photometric band, respectively. 

As a final point, let us mention that 
the assumption of a canonical relation
between mass and luminosity plays only a marginal role 
on the pulsational 
scenario discussed in the present investigation. 
This is shown in Fig. 22 
and Fig. 23 where we compare 
the predicted PL$_V$ and PL$_K$ relations inferred from 
canonical stellar evolution with the pulsating models based 
on evolutionary computations which account for a mild convective 
core overshooting (Chiosi et al. 1992, 1993). We notice that the 
overshooting models would give rather similar relations, at least 
with log$P\le$ 1.8, thus suggesting that 
the adopted canonical luminosities (vanishing overshooting) play a 
minor role in determining the pulsational behavior of Cepheids. 
However, the results in Fig. 22 somehow disagree with the clear 
separation between fundamental and first
overtone pulsators observed by the microlensing surveys, thus providing    
some support to the assumption of 
canonical luminosities (see Fig. 13). Note that, since canonical models 
with mass loss are simulating overshooting models without mass loss, 
such a suggestion gives also constraints on the amount 
of mass loss preceding He--burning evolution. 

{\it Acknowledgements.} We want to thank A. Weiss for discussion and 
useful comments. Thanks also to the anonymous referee who helped us to 
improve the final version of the paper.  
This work was partially supported by CNAA, 
through a postdoc research grant to M. Marconi, ASI and CRA. 

\pagebreak

\pagebreak

\figcaption [] {Distribution in the log$T_e$-log$P$ plane of fundamental
(F) and first-overtone (FO) 
pulsators with different masses and 
chemical compositions. For 
each given mass and chemical composition the luminosity 
level has been fixed according to a canonical mass-luminosity relation 
(see text).} 

\figcaption [] {Location in the log$P$-$<M_{bol}>$ plane of fundamental 
pulsators. In each panel the solid line refers to the quadratic least squares 
fit of theoretical models at the labelled metallicity, whereas the
dashed line, in the upper two panel, shows the solar metallicity relation.} 

\figcaption [] {Location in the log$P$-$<M_{V}>$ plane of fundamental 
and first--overtone 
pulsators with 
different masses and chemical compositions, in 
comparison with current empirical PL$_V$ 
relations (see text). Symbols as in Fig. 1.} 

\figcaption [] {{\it (lower panel}) -- Comparison in the 
log$P$-$<M_V>$ plane of fundamental pulsators with $Y$=0.25, 
$Z$=0.004 (filled squares) and SMC Cepheids (open circles: data 
by Laney \& Stobie 1994 [LS94]) by adopting a distance modulus 
$DM_{SMC}$=19.1. The solid lines show the predicted blue 
and red edges of the instability strip with $Z$=0.004. 
{\it (upper panel}) -- As in the 
lower panel, but for models 
with $Y$=0.25, $Z$=0.008 (filled triangles) and LMC Cepheids 
with $DM_{LMC}$=18.6. In this case the solid lines show the 
predicted blue 
and red edges of the instability strip with $Z$=0.008, whereas the 
dashed lines refer to $Z$=0.004.}

\figcaption [] {Comparison in the log$P$-$<M_V>$ plane of fundamental
pulsators with $Y$=0.28, $Z$=0.02 (filled circles) and 
Galactic Cepheids (open circles: data by LS94 and Gieren, Fouqu\`e \& 
Gomez 1997 [GFG97]). 
The solid lines
show the predicted blue and red
edges of the instability strip with $Z$=0.02, 
while the dashed lines refer to $Z$=0.008.}

\figcaption [] {As in Fig. 3, but for infrared magnitudes.}

\figcaption [] {As in Fig. 4, but for infrared magnitudes.}

\figcaption [] {As in Fig. 5, but for infrared magnitudes.}

\figcaption [] {Comparison between theoretical models (symbols as
in Fig. 1) and empirical log$P$--($B-V$) relations from CC86.}

\figcaption [] {Comparison between theoretical models (symbols as
in Fig. 1) and empirical log$P$--($B-V$) (lower panel)
and log$P$--($V-K$) (upper panel) relations from LS94.}

\figcaption [] {Comparison between LS94 data for Galactic
and Magellanic Cepheids in the log$P$--($B-V$)
plane and the predicted
edges of the instability strip at the three labeled metallicities.}

\figcaption [] {As in Fig. 11, but in the log$P$--($V-K$) plane.}

\figcaption [] {Predicted PL$_V$ relations for fundamental 
and first--overtone pulsators with different 
chemical compositions (see Eq. 1a--1d). Symbols as in Fig. 1.}

\figcaption [] {As in Fig. 10, but showing the linear approximation 
for fundamental pulsators with periods longer and shorter than log$P\sim 1.4$.} 

\figcaption [] {Predicted PL$_K$ relations for fundamental and 
first--overtone pulsators with different 
chemical compositions (see Eq. 2a--2d). Symbols as in Fig. 1.}

\figcaption [] {Predicted (projected onto a plane) optical PLC relations 
for fundamental pulsators with different chemical 
compositions (see Table 1). Symbols as in Fig. 1.}

\figcaption [] {As in Fig. 16, but for the infrared PLC relations.}

\figcaption [] {Observed data (open circles) 
and empirical PL$_V$ 
relations (dashed line) for Galactic Cepheids 
(data by LS94 and GFG97) in comparison with  
predicted relations at $Z$=0.02 (solid line) and $Z$=0.008 (dotted line).}

\figcaption [] {Observed data (open circles) 
and empirical PL$_K$ 
relations (dashed line) for Galactic Cepheids 
(data by LS94 and GFG97) in comparison with the
predicted relation at $Z$=0.02 (solid line).}

\figcaption [] {The theoretical projected infrared PLC relation 
for $Z$=0.02 (solid line) compared with the location of 
Galactic Cepheids (filled points: data by LS94 and GFG97).}

\figcaption [] {Individual distance modulus of LMC Cepheids 
(filled points), with the related error bar, as obtained from LS94 data 
and predicted PL and 
PLC relations with different metallicities. The dashed line is a reference 
level of 18.7 mag.}

\figcaption [] {As in Fig. 13, but by adopting overshooting models.
The theoretical relations are the same as in Fig. 13.}

\figcaption [] {As in Fig. 15, but by adopting overshooting models.
The theoretical relations are the same as in Fig. 15.}


\vspace{5.0mm} 
\begin{deluxetable}{ccccc}
\tablecaption{Theoretical PLC relations for
fundamental pulsators.}\label{tbl-1} 
\tablehead{ 
\colhead{Z\tablenotemark{a}} & 
\colhead{$\alpha$\tablenotemark{b}} & 
\colhead{$\beta$\tablenotemark{c}} &
\colhead{$\gamma$\tablenotemark{d}} & 
\colhead{$\sigma$\tablenotemark{e}} } 
\startdata 
  \multicolumn{5}{c}{$<M_V>$=$\alpha$+$\beta$log$P$+$\gamma$($<B>-<V>$)} \nl  
      &         &                &         &         \nl 
0.004 &-2.54 & -3.52 & 2.79 & 0.04 \nl
      &$\pm$0.04\tablenotemark{f}&$\pm$0.03&$\pm$0.07& \nl
0.008 &-2.63 & -3.55 & 2.83 & 0.04 \nl
      &$\pm$0.04&$\pm$0.03&$\pm$0.06& \nl
0.02  &-2.98 & -3.72 & 3.27 & 0.08 \nl
      &$\pm$0.07&$\pm$0.10&$\pm$0.18& \nl
      &          &         &         &         \nl 
    \multicolumn{5}{c}{$<M_K>$=$\alpha$+$\beta$log$P$+$\gamma$($<V>-<K>$)} \nl 
         &         &         &         &         \nl 
0.004 &-3.44 & -3.61 & 0.64 & 0.03 \nl
      &$\pm$0.04&$\pm$0.03&$\pm$0.03& \nl
0.008 &-3.37 & -3.60 & 0.61 & 0.04 \nl
      &$\pm$0.04&$\pm$0.03&$\pm$0.03& \nl
0.02  &-3.25 & -3.55 & 0.53 & 0.04 \nl
      &$\pm$0.04&$\pm$0.05&$\pm$0.04& \nl
\enddata
\tablenotetext{a}{Metal content.
\hspace*{0.5mm} $^b$ Zero point.
\hspace*{0.5mm} $^c$ Logarithmic period coefficient.
\hspace*{0.5mm} $^d$ Color coefficient.  
\hspace*{0.5mm} $^e$ Standard deviation (mag).
\hspace*{0.5mm} $^f$ Errors on the coefficients.}
\end{deluxetable}

                             
\begin{deluxetable}{ccc}
\tablecaption{SMC and LMC distance moduli derived by applying 
theoretical PL and PLC relations to the reddening-corrected
magnitudes and colours by LS94.}\label{tbl-1}
\tablehead{
\colhead{Method\tablenotemark{a}} &
\colhead{$DM_{SMC}(Z=0.004)$\tablenotemark{b}} &
\colhead{$DM_{LMC}(Z=0.008)$\tablenotemark{c}} }
\startdata
                      &         &         \nl
PL($<M_V>$) &18.98$\pm$0.28\tablenotemark{d} & 18.42$\pm$0.30 \nl
PL($<M_K>$) &19.14$\pm$0.15 & 18.62$\pm$0.12 \nl
PLC($<M_V>$,$<B>-<V>$  &19.19$\pm$0.17& 18.62$\pm$0.17 \nl
PLC($<M_K>$,$<V>-<K>$  &19.28$\pm${0.17}& 18.74$\pm${0.13} \nl
\enddata
\tablenotetext{a}{Method adopted to estimate the distance modulus.
\hspace*{0.5mm} $^b$ SMC distance modulus (mag).
\hspace*{0.5mm} $^c$ LMC distance modulus (mag).
\hspace*{0.5mm} $^d$ Standard deviation (mag).}
\end{deluxetable}

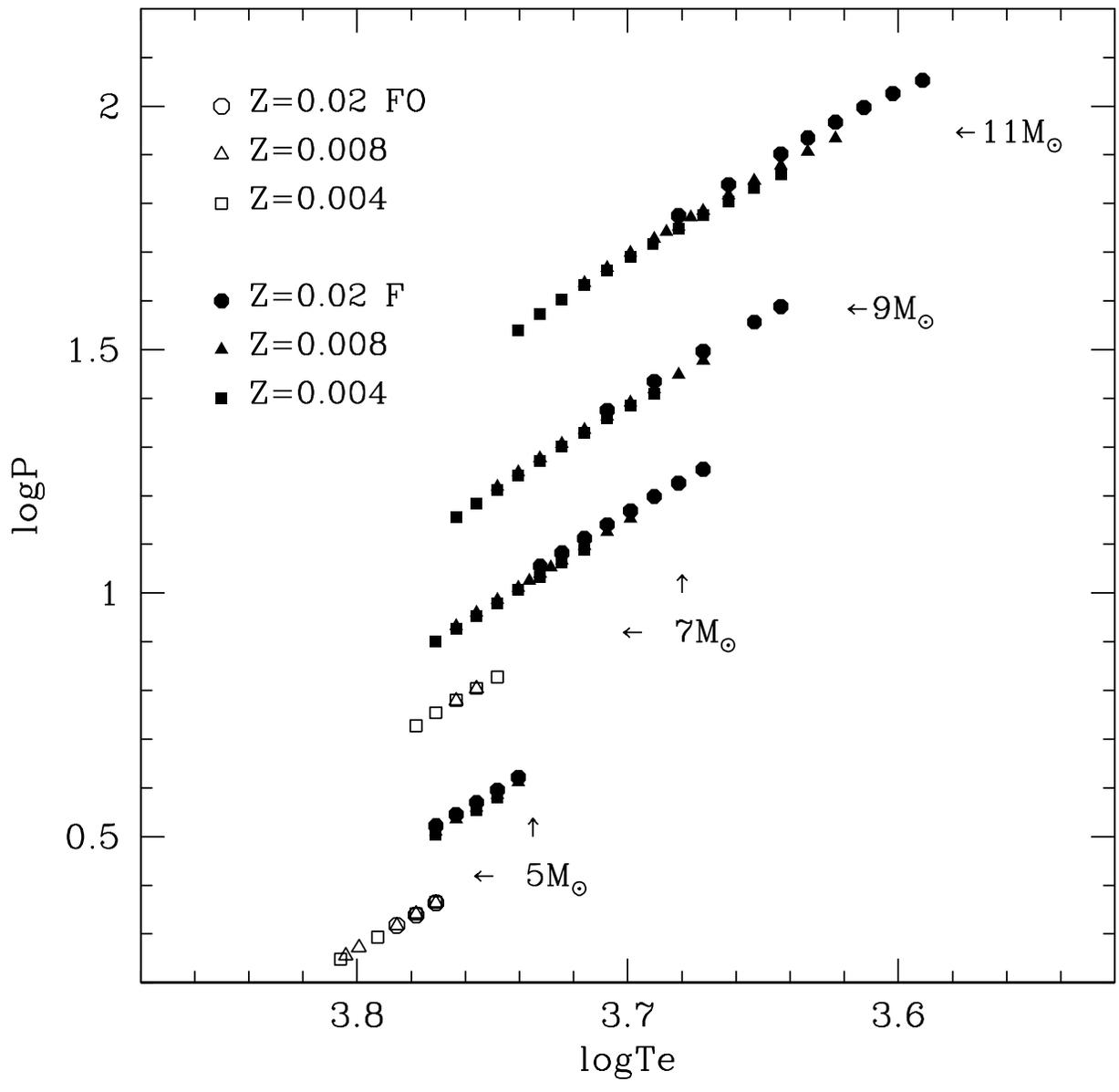
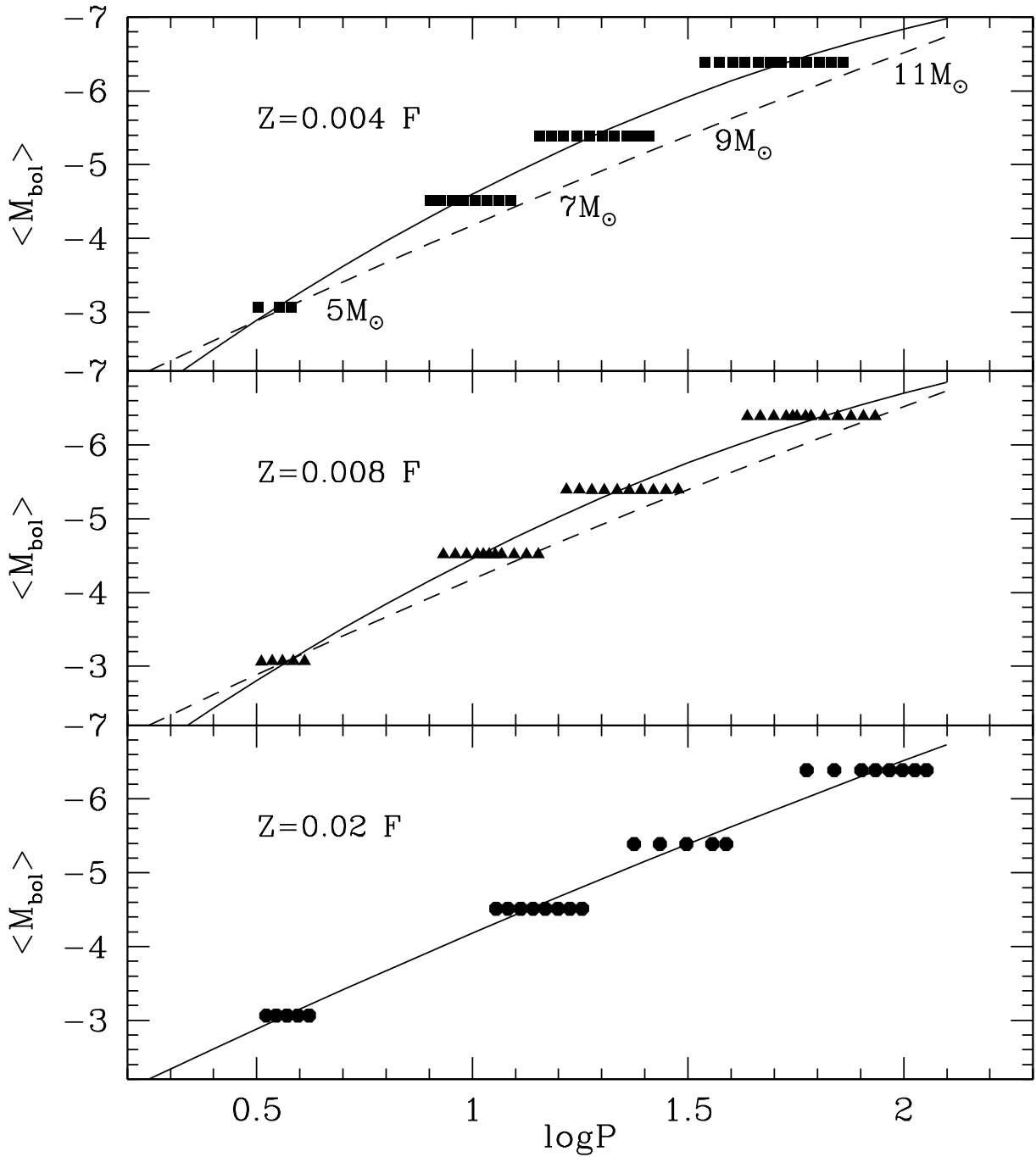
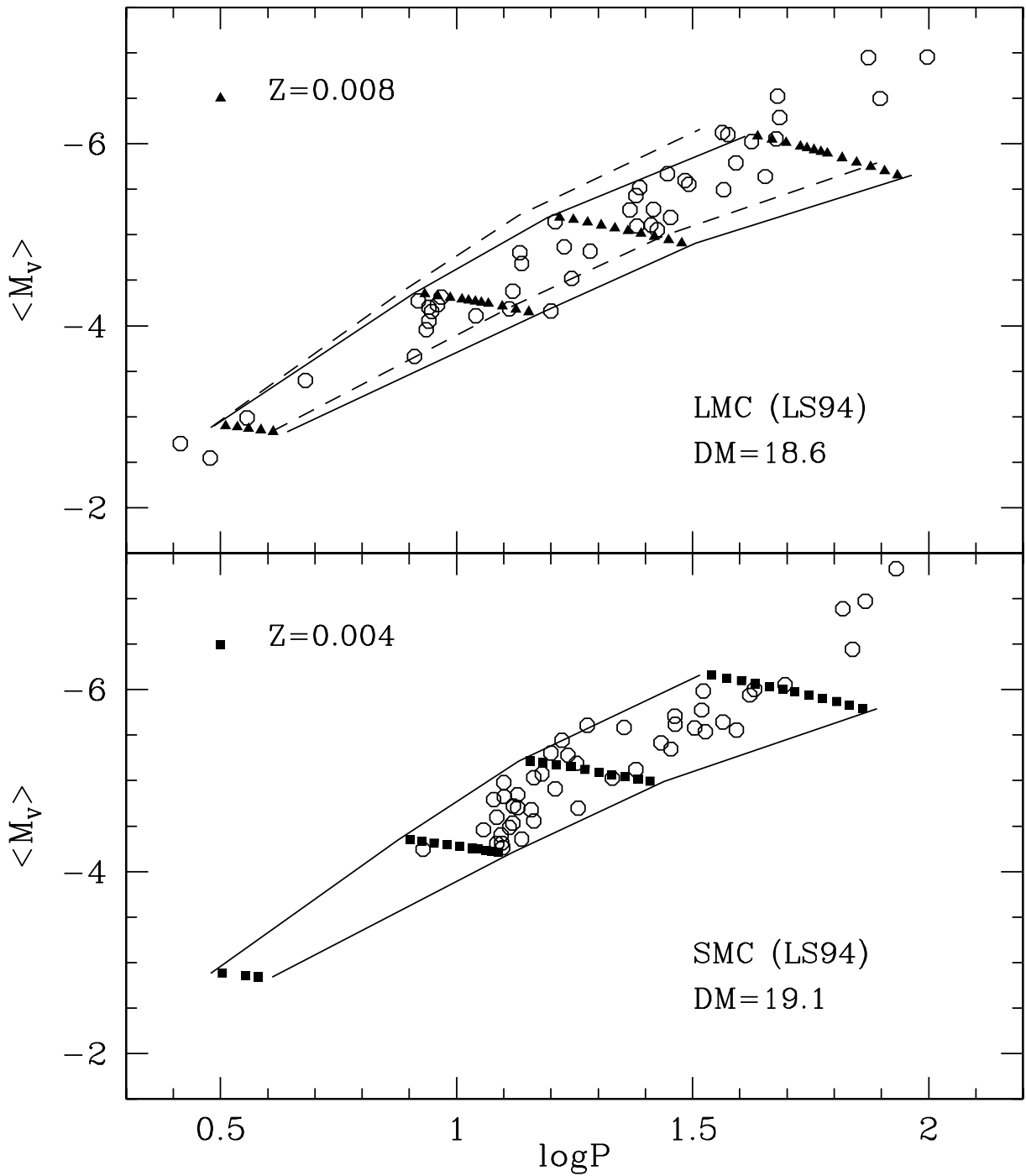
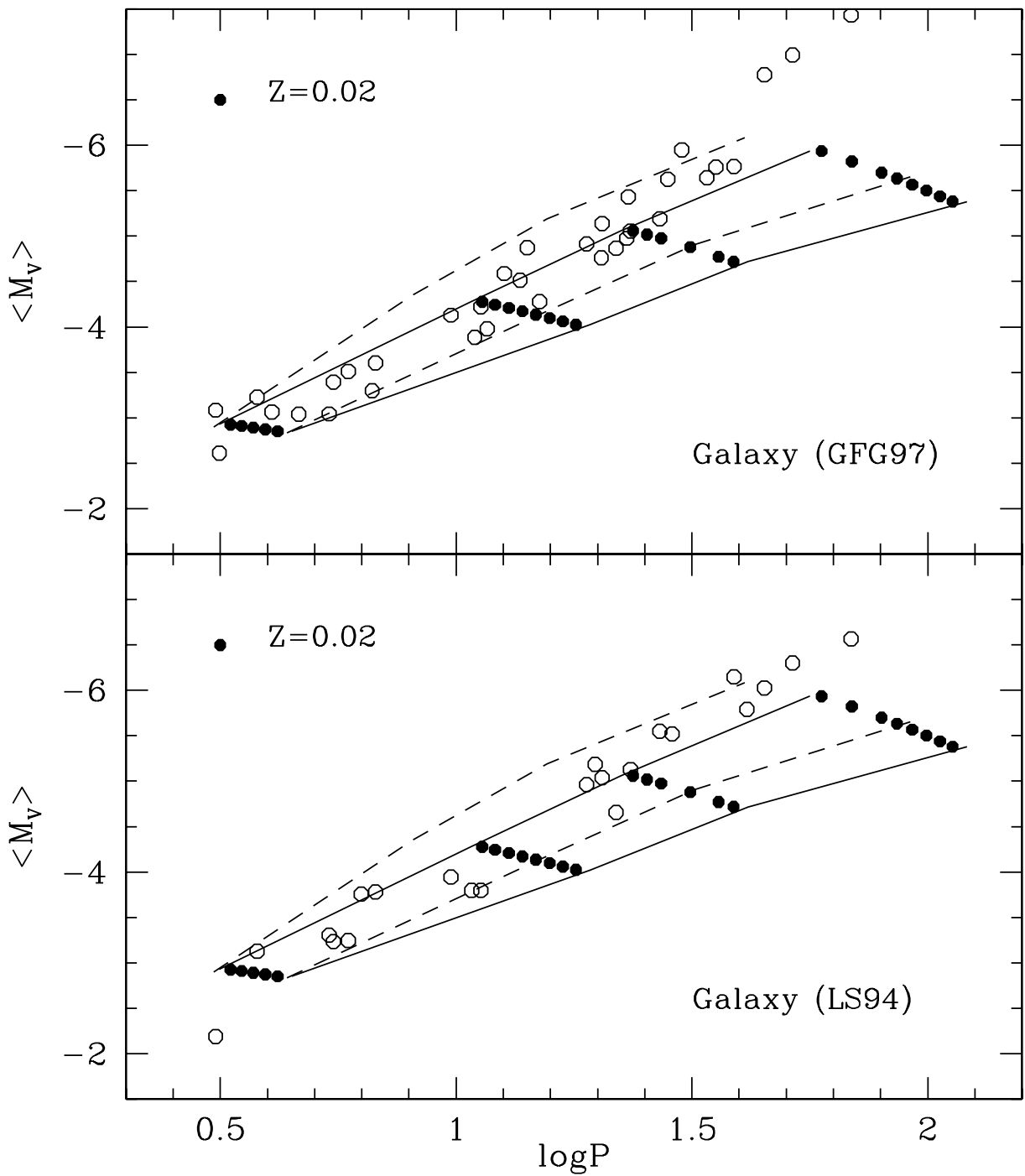
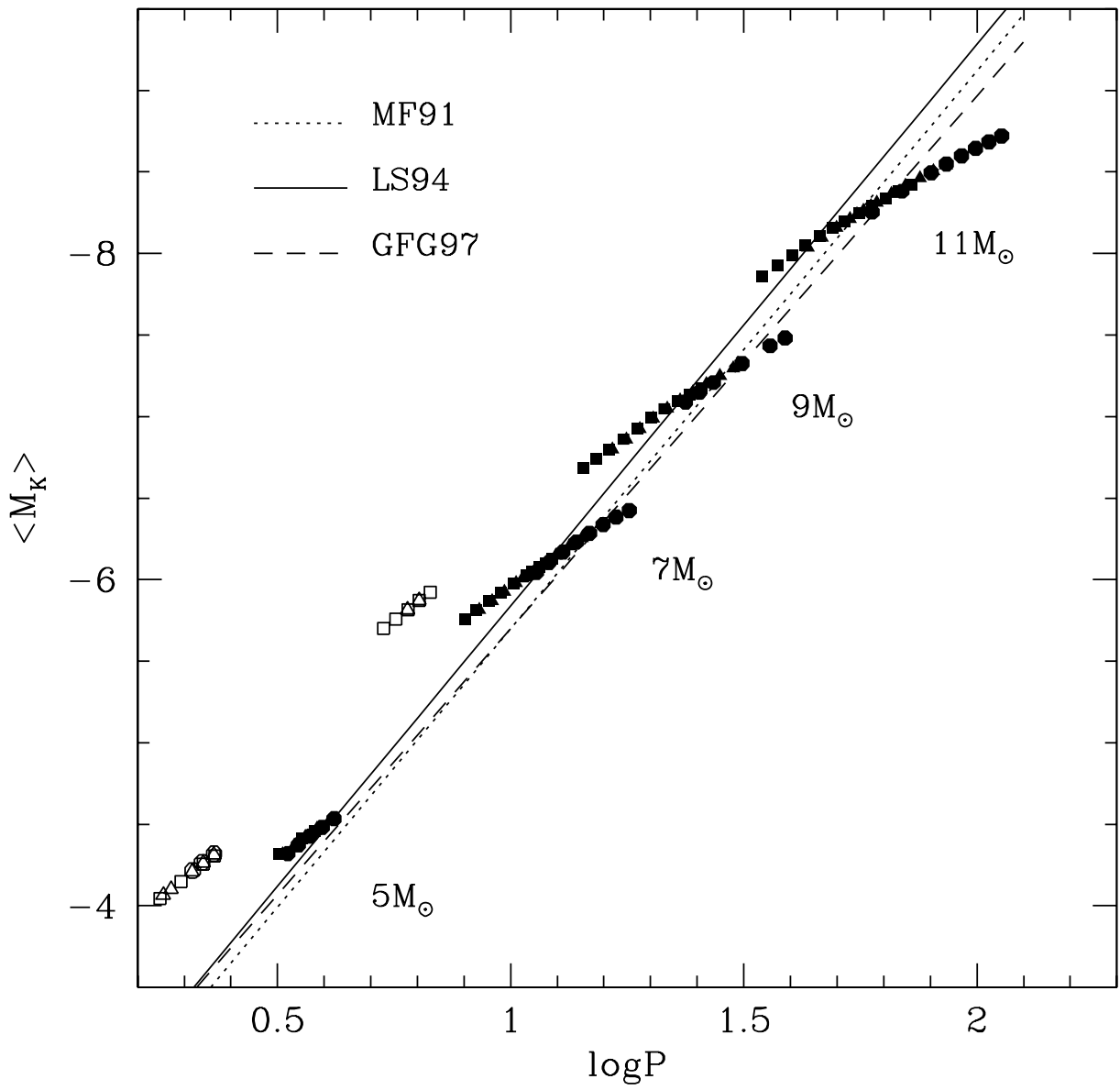
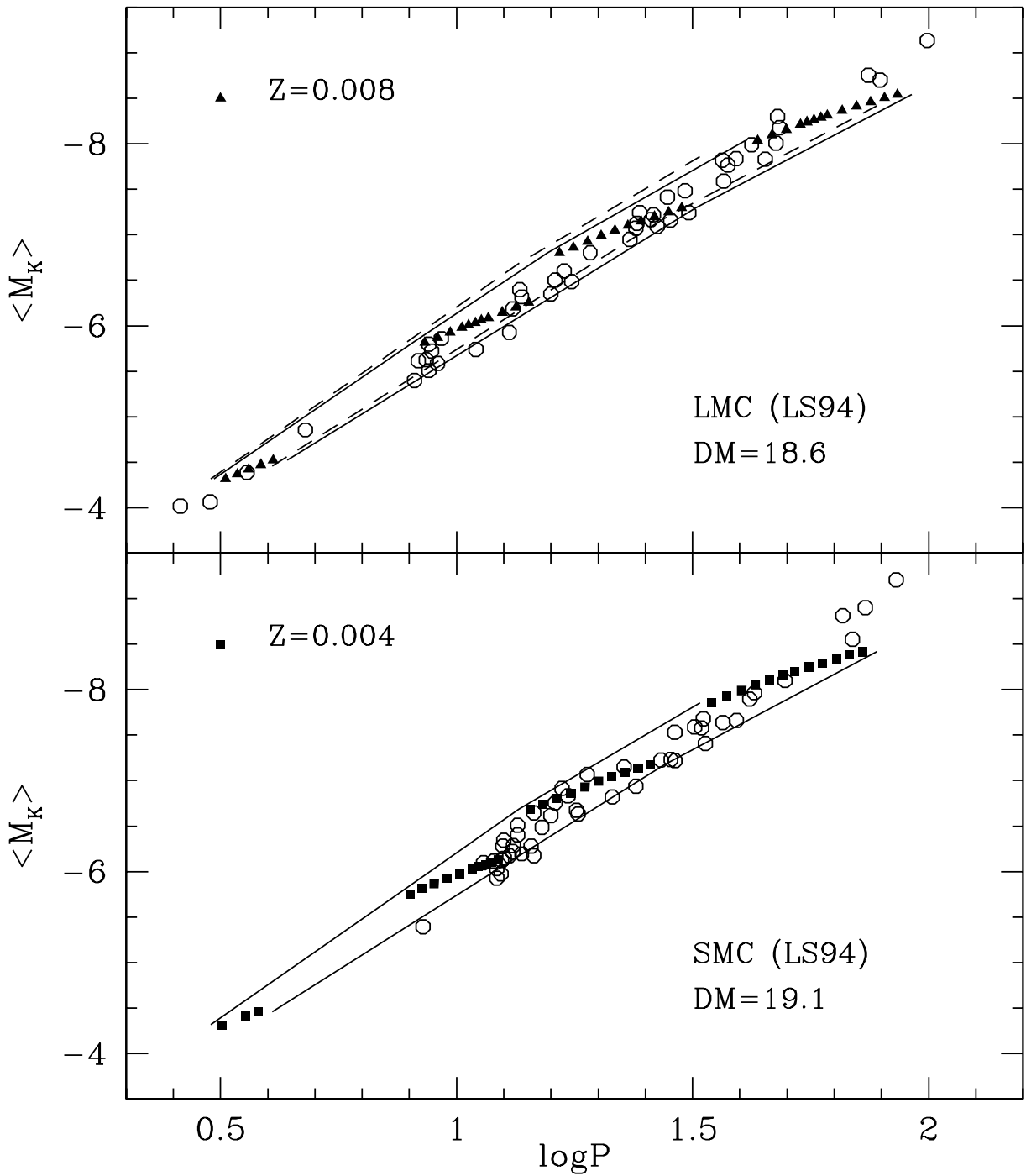
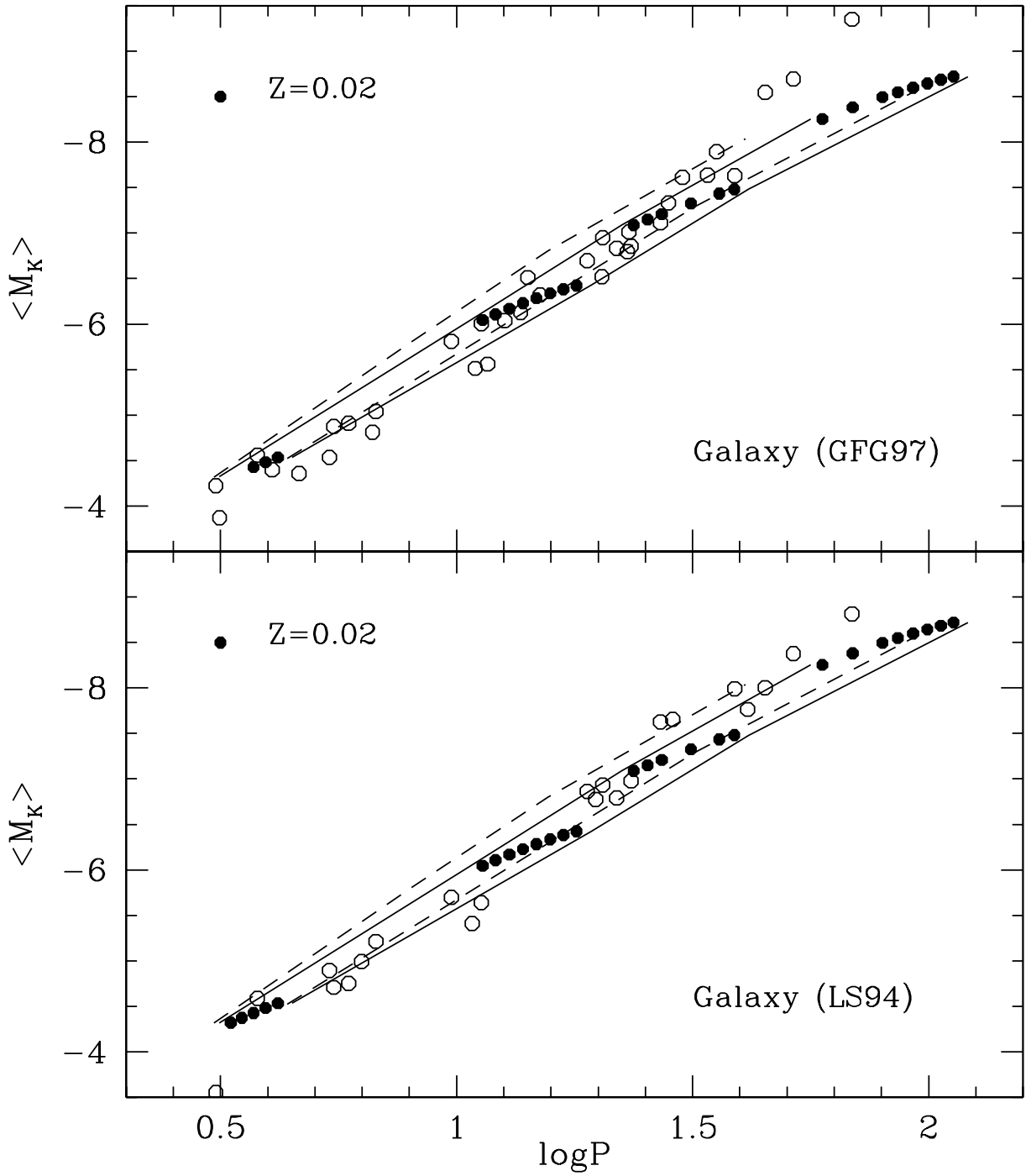
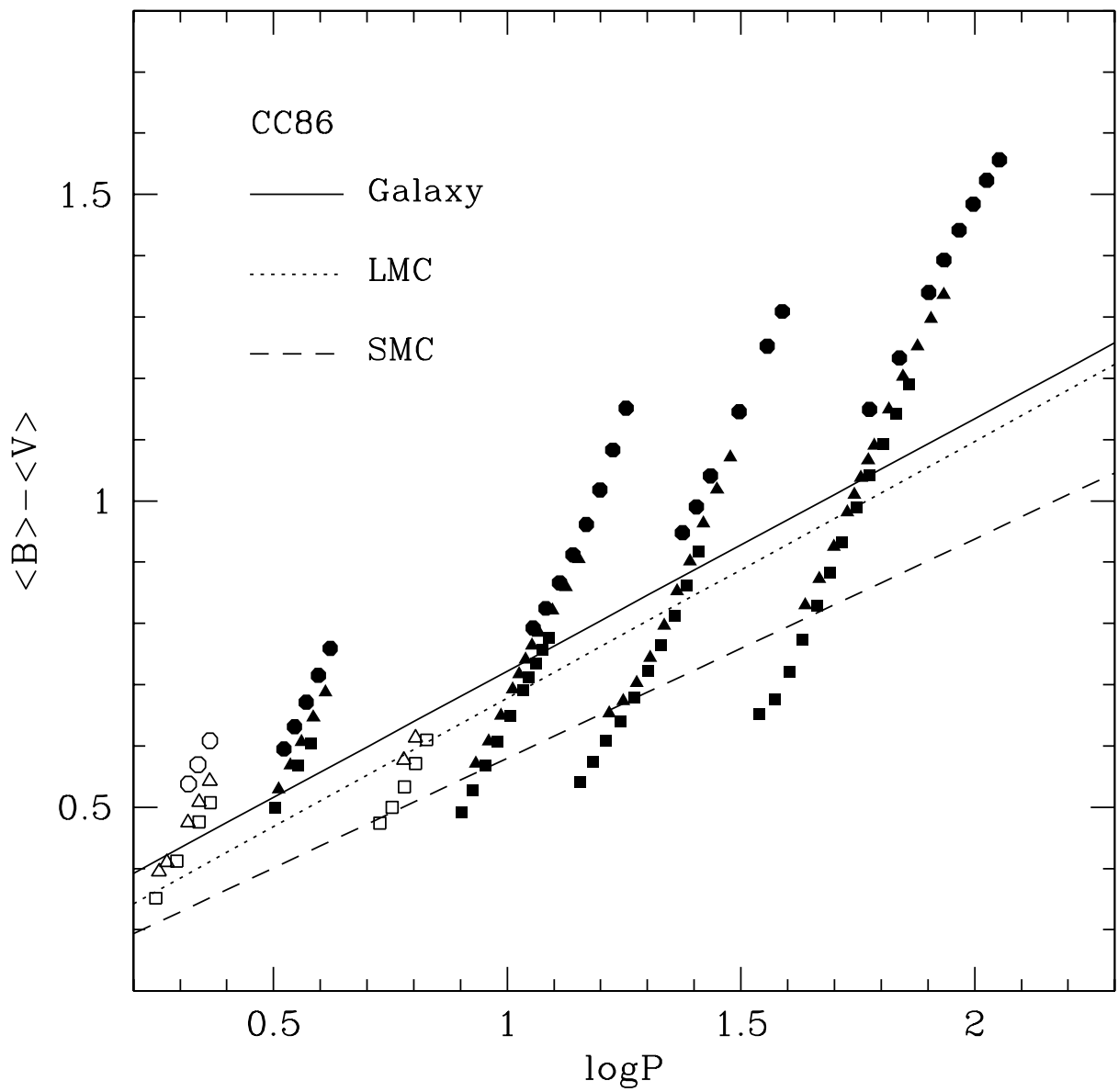
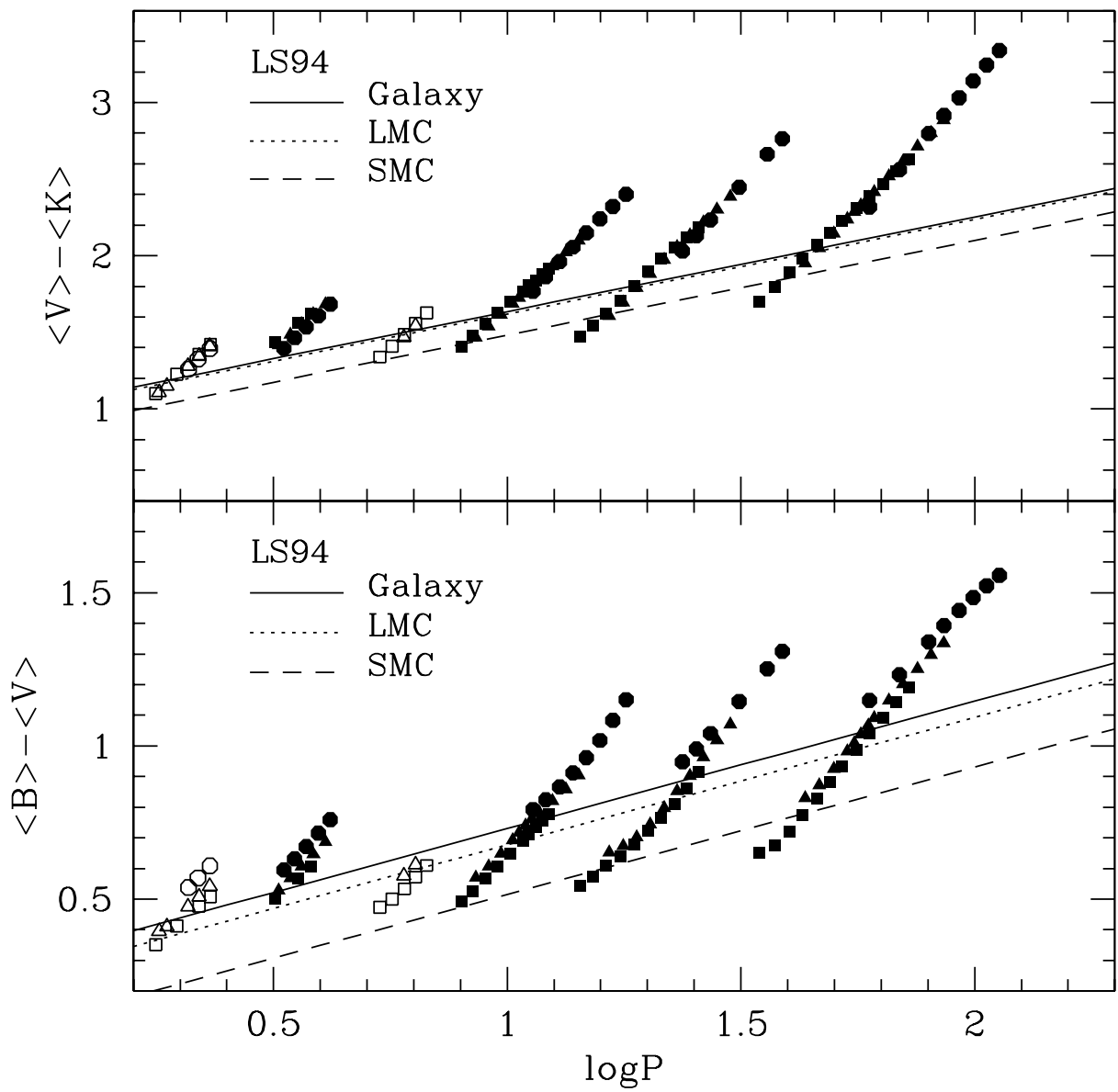
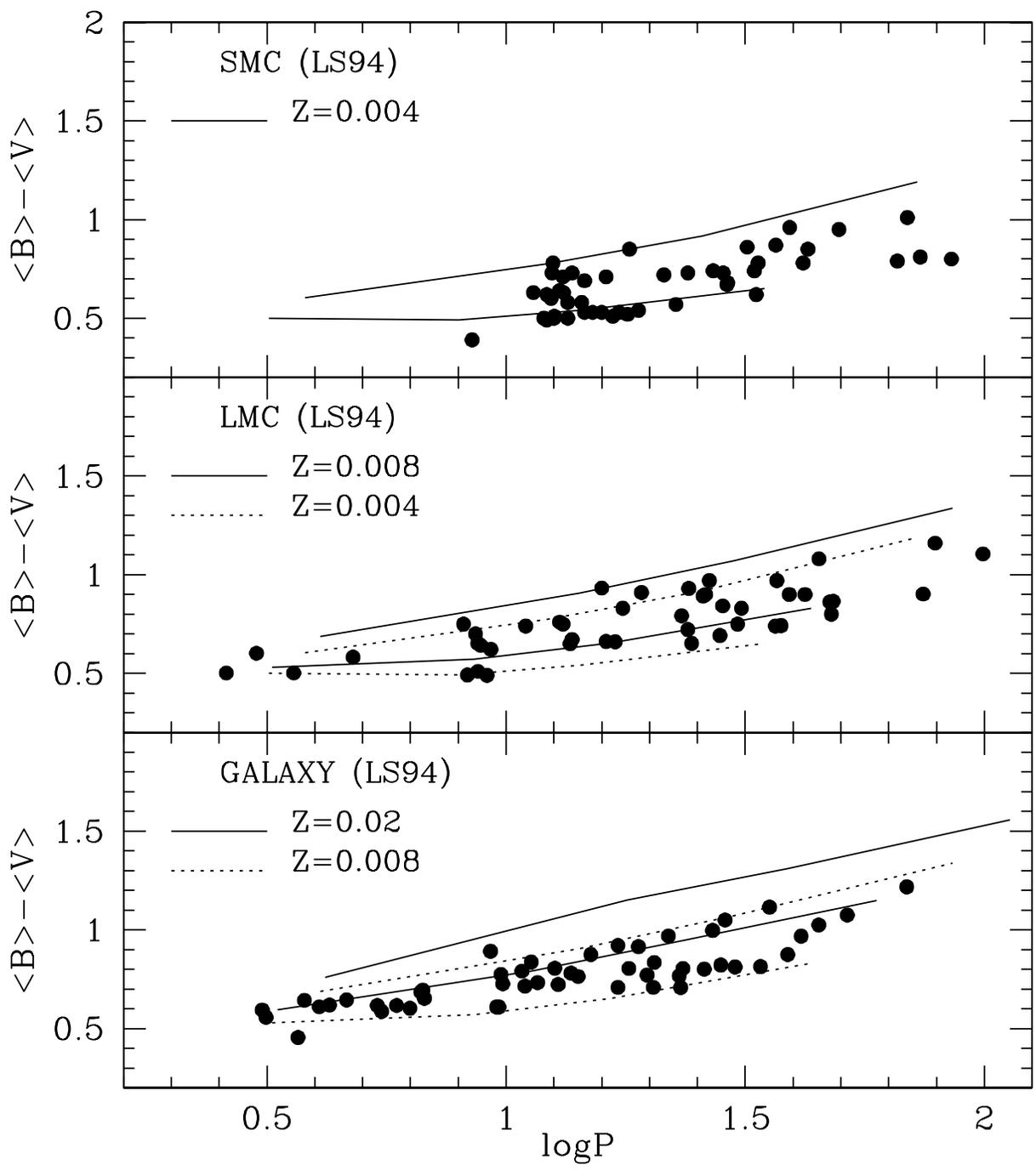
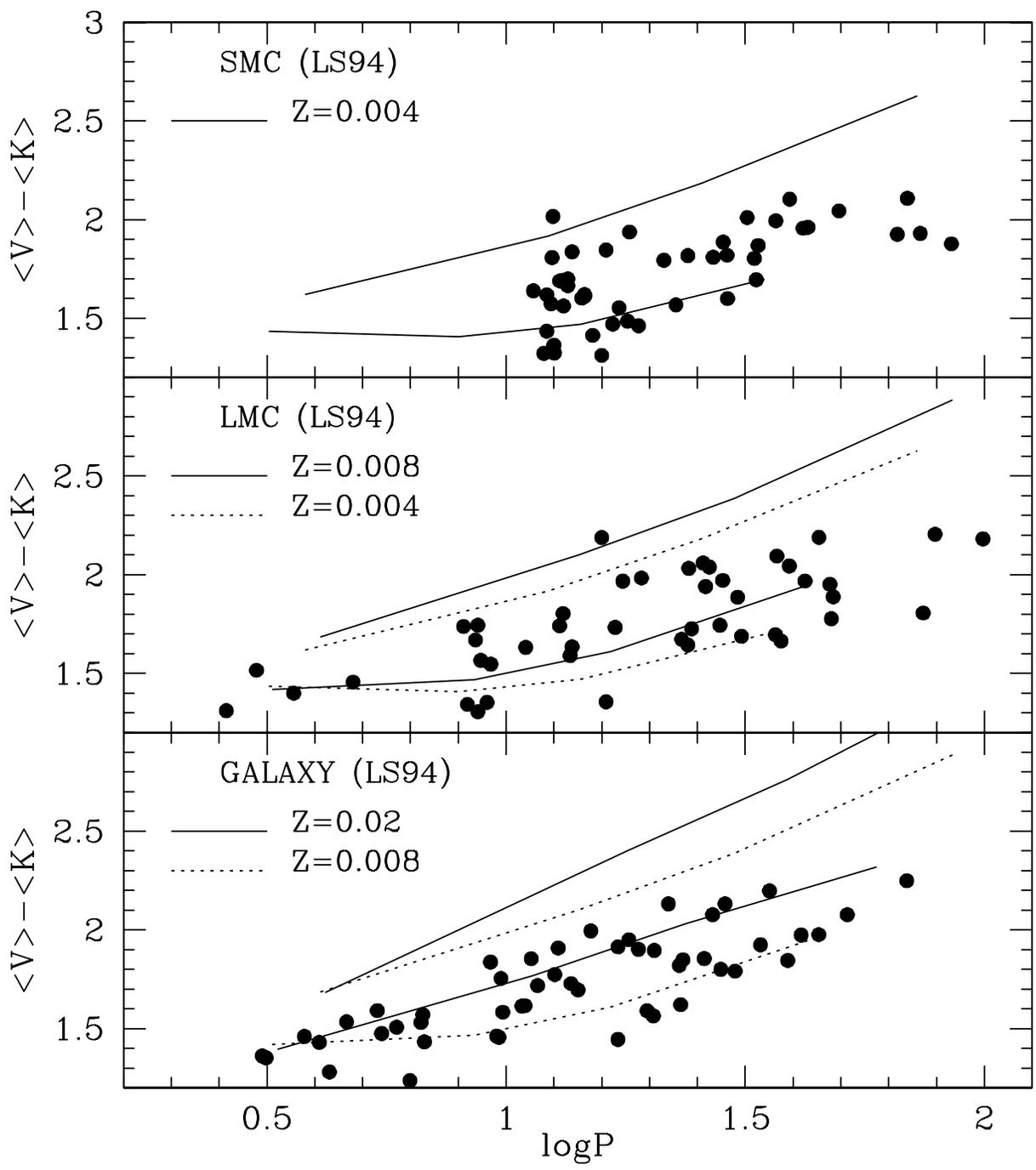
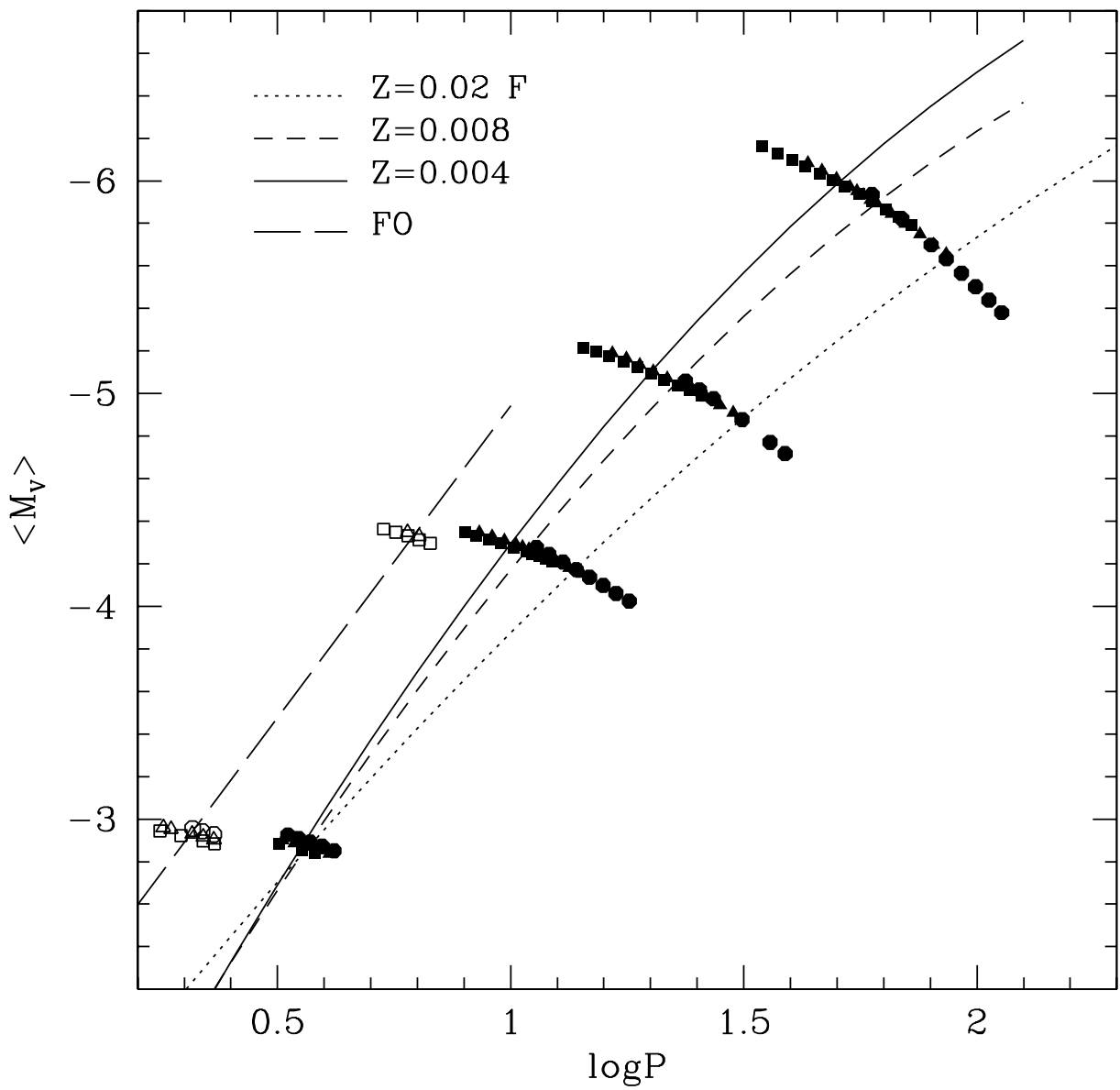
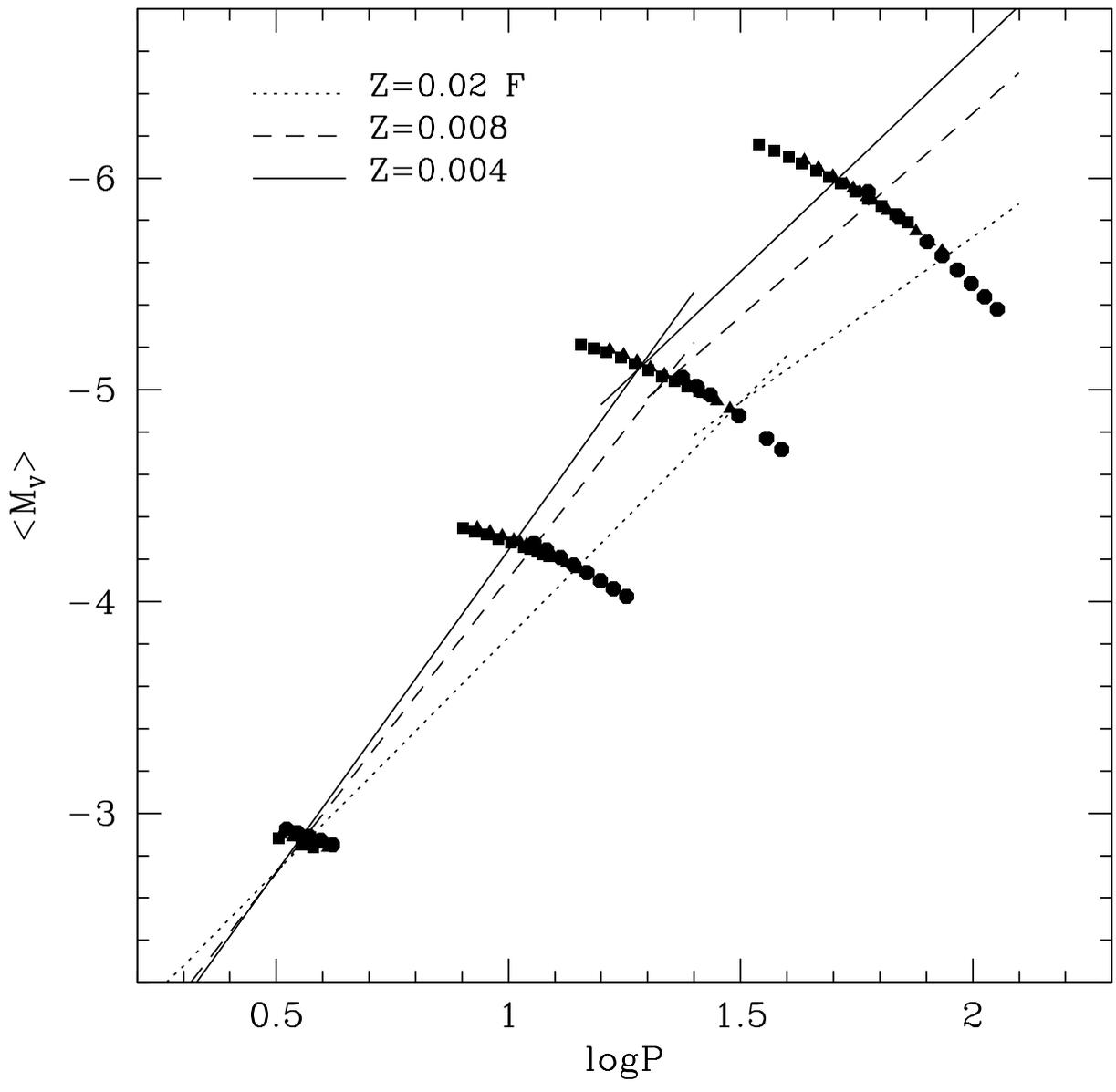
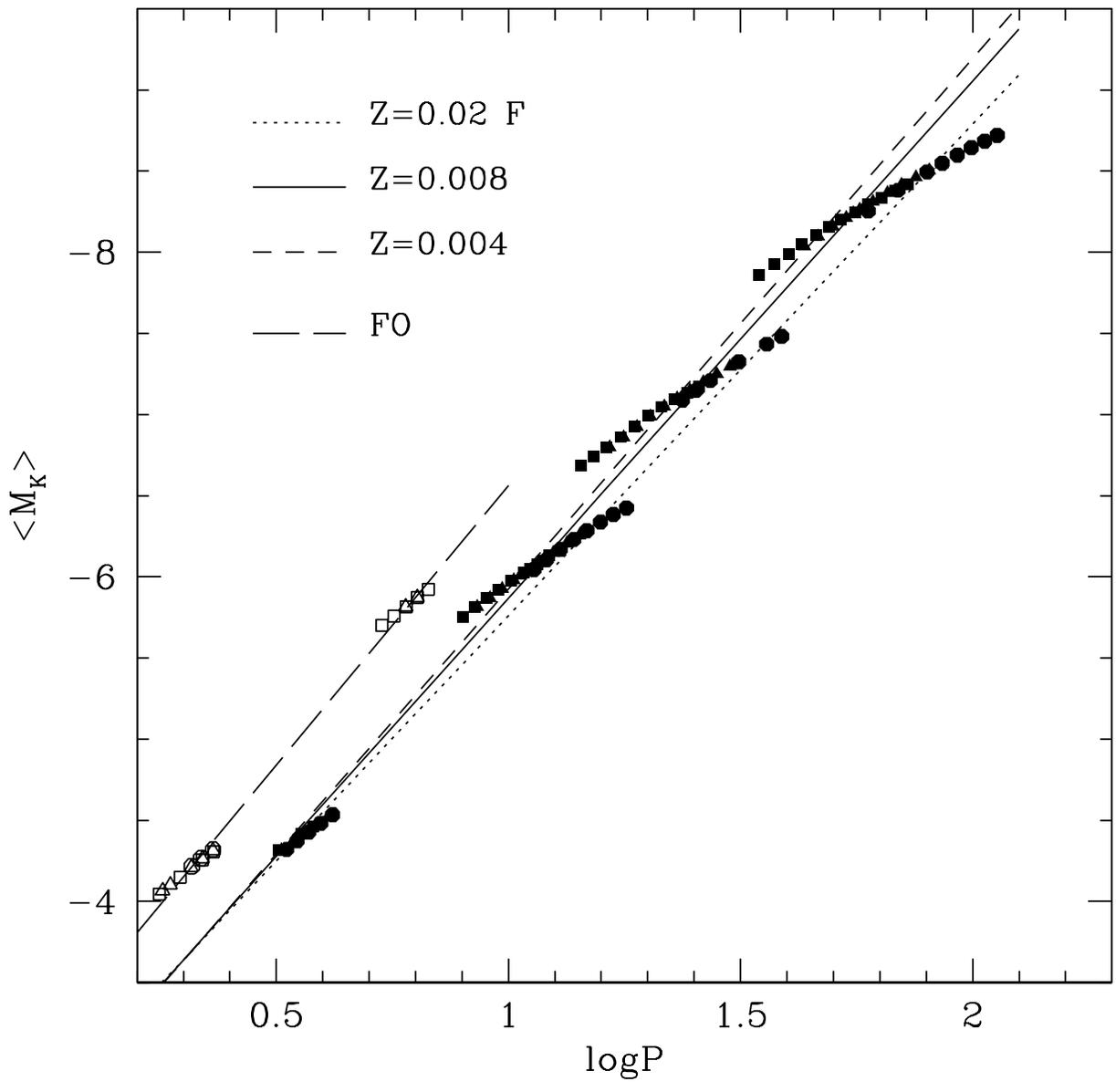
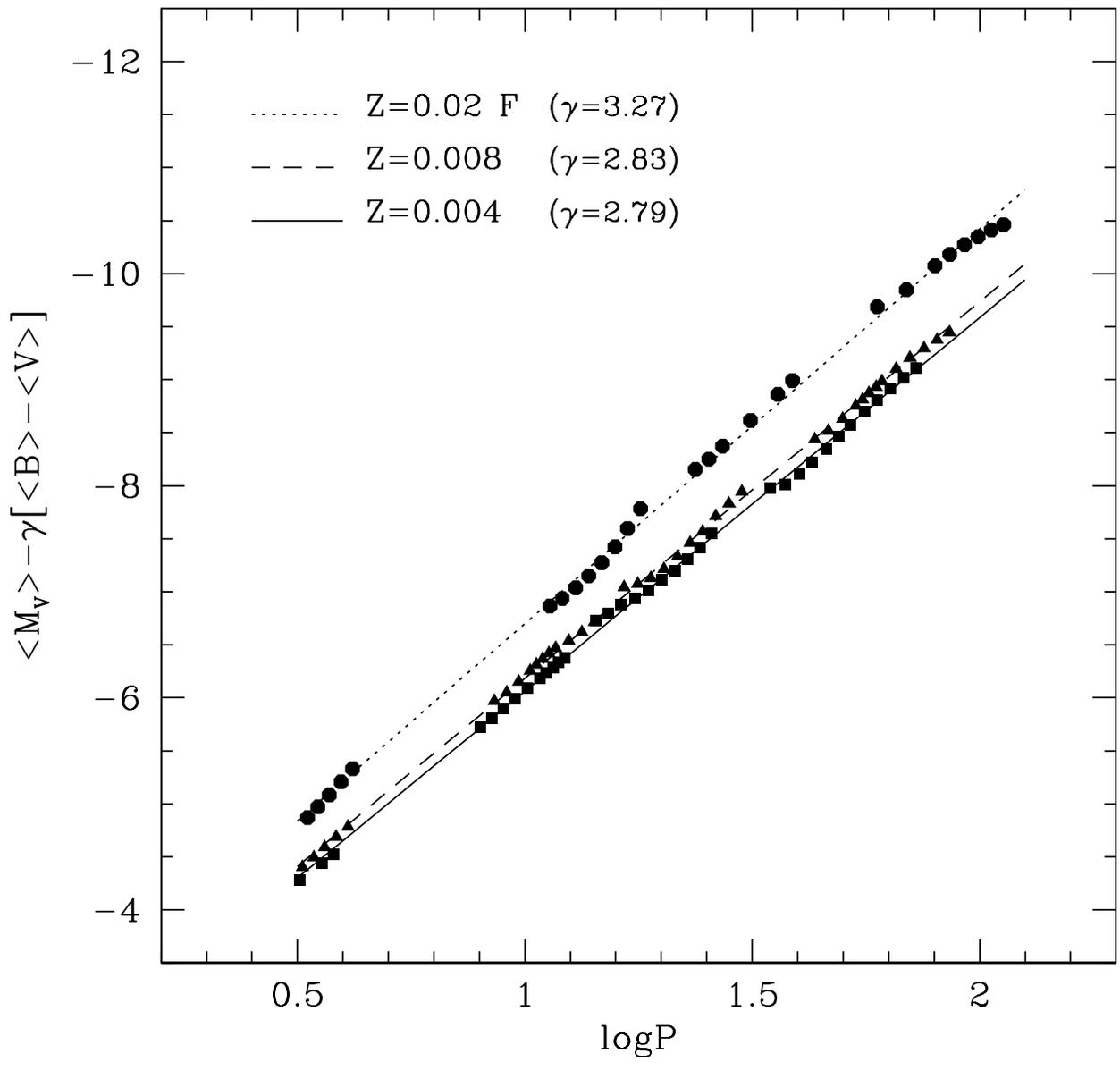
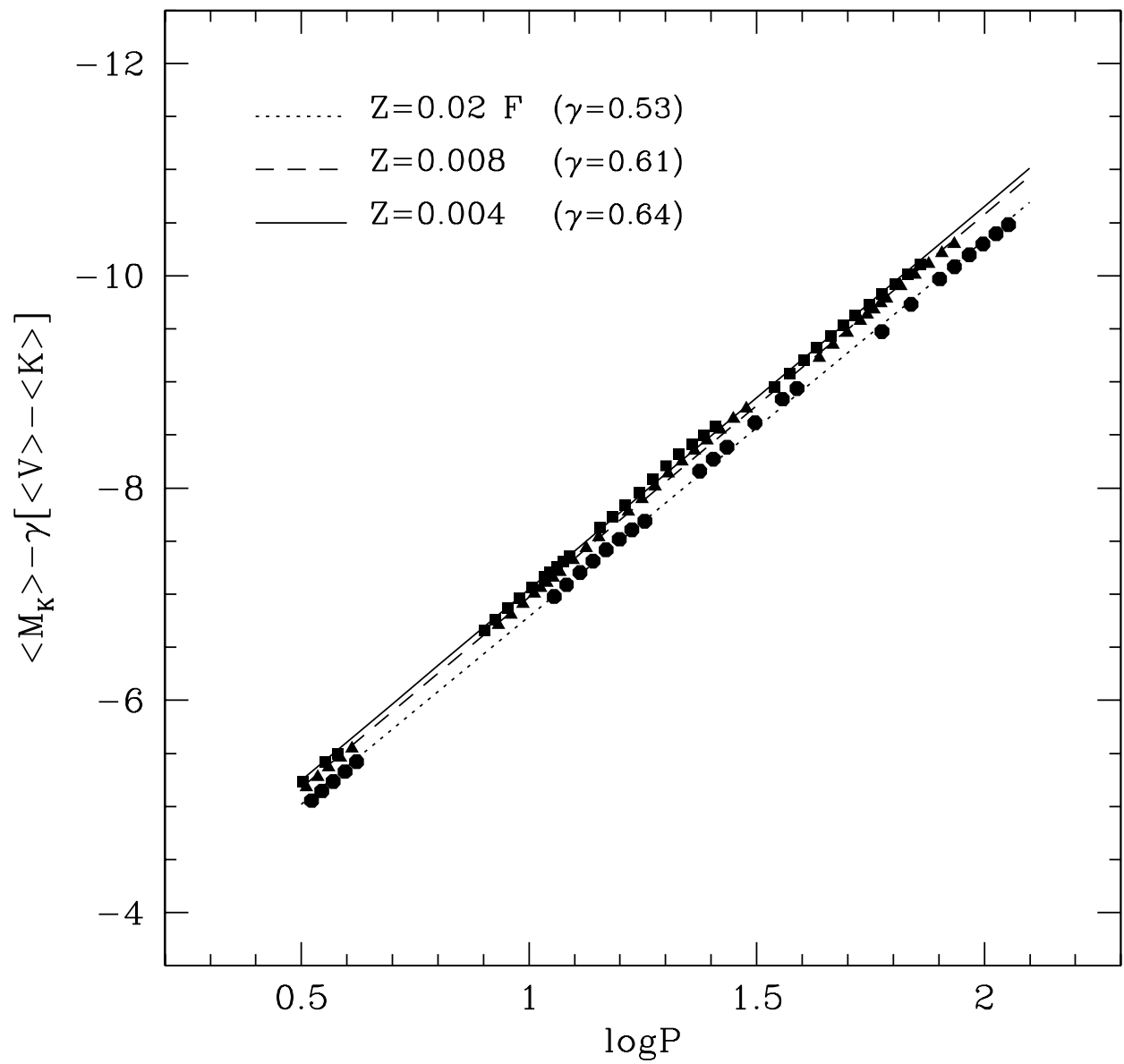
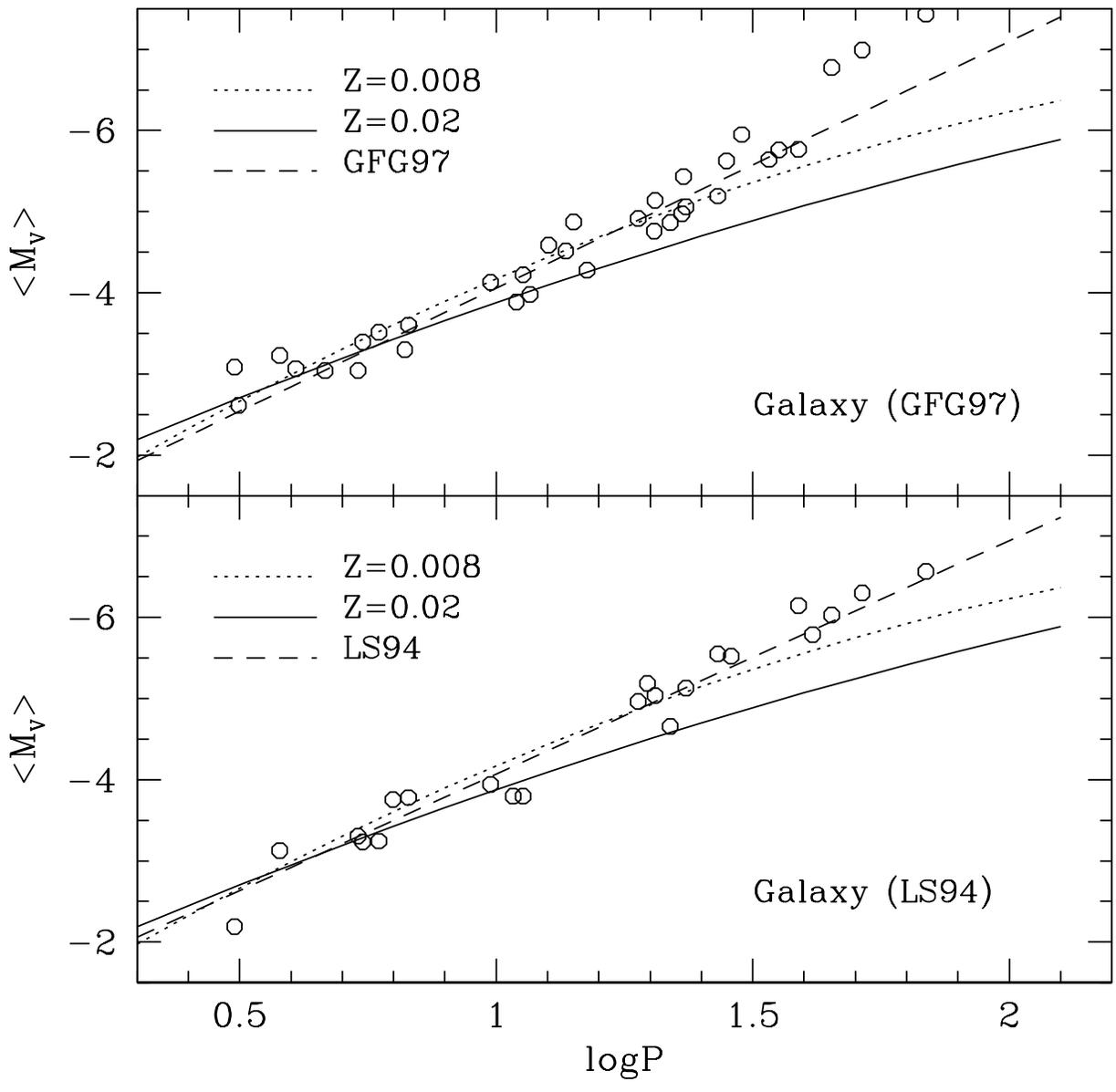
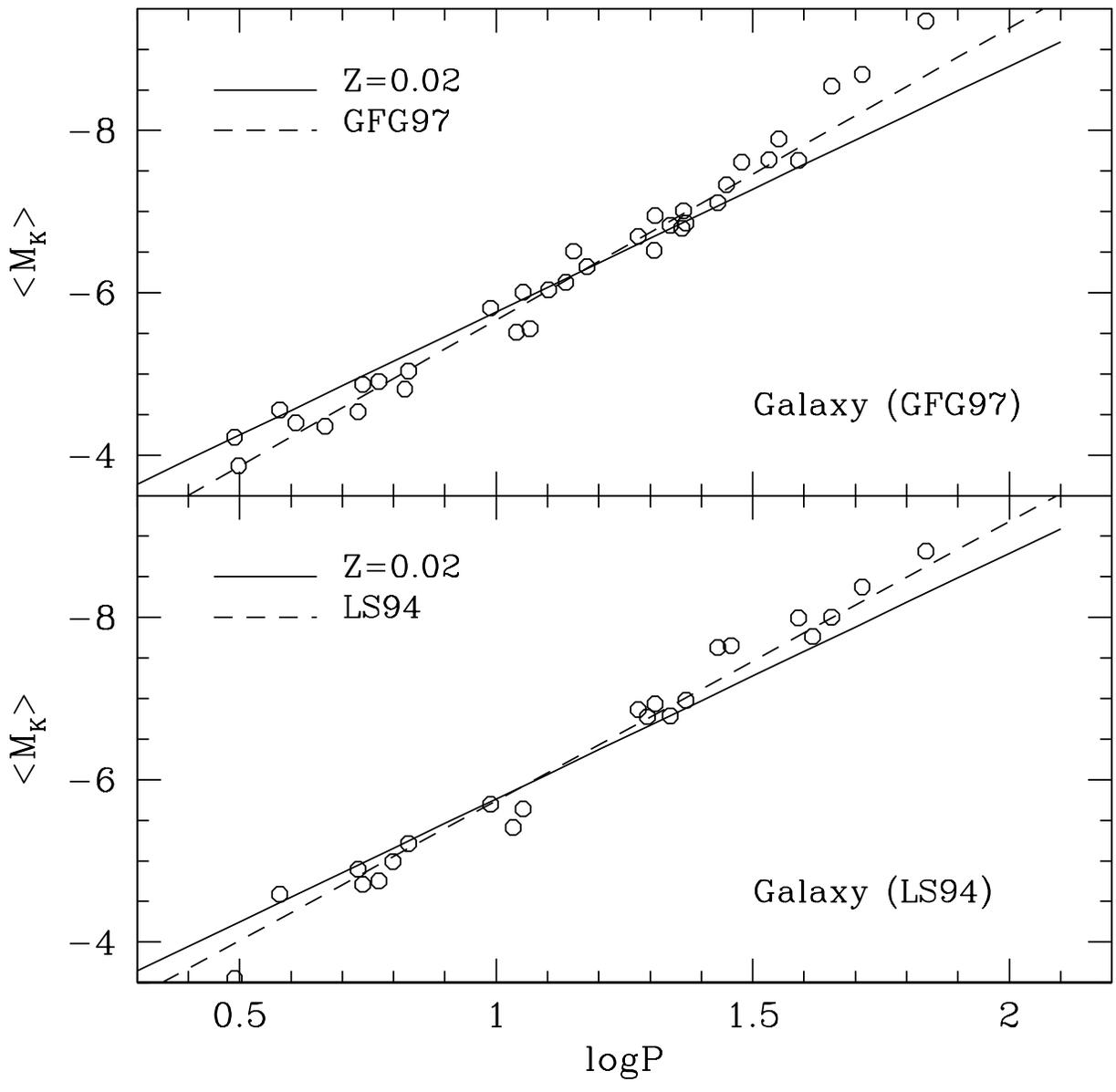
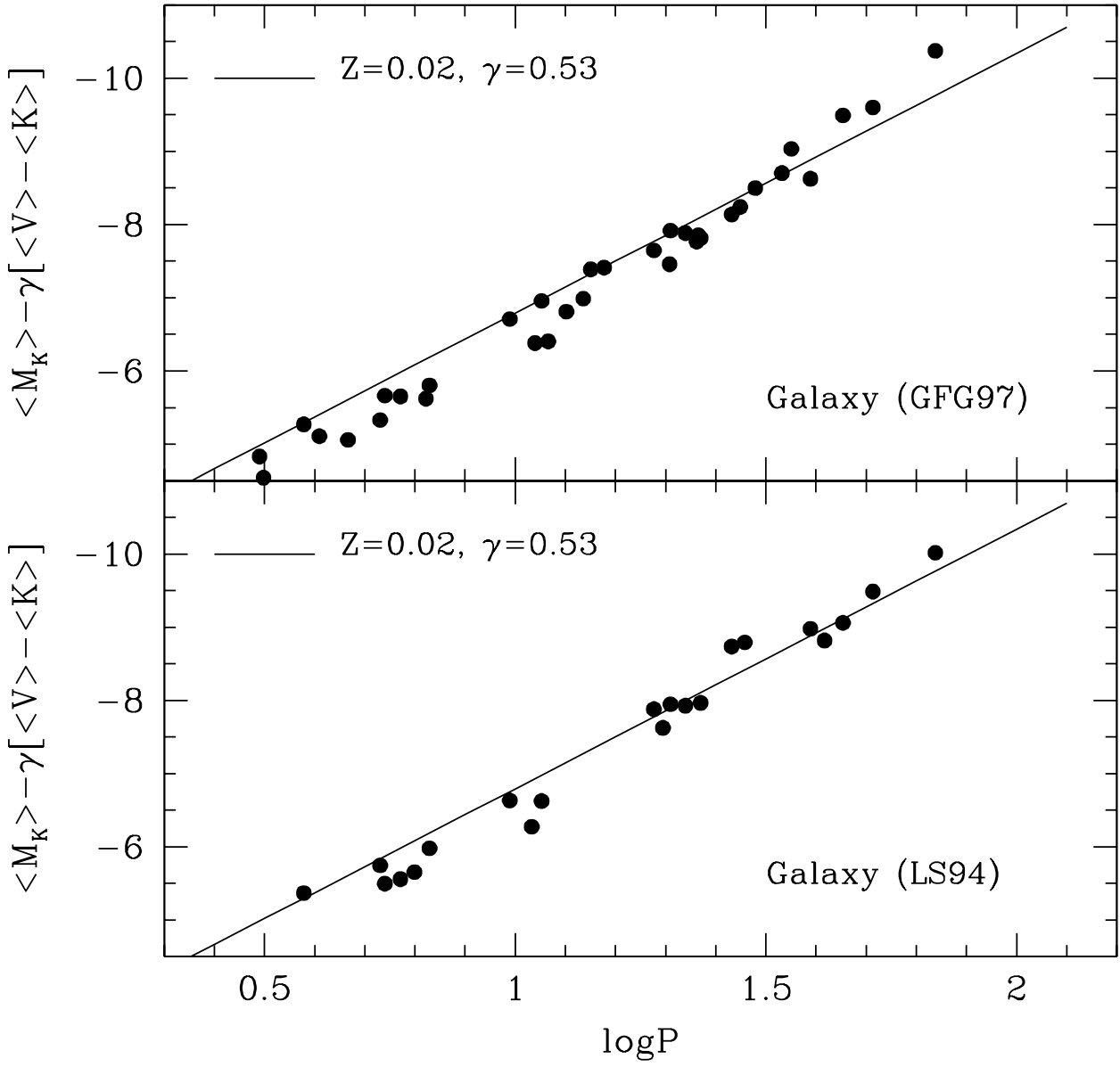
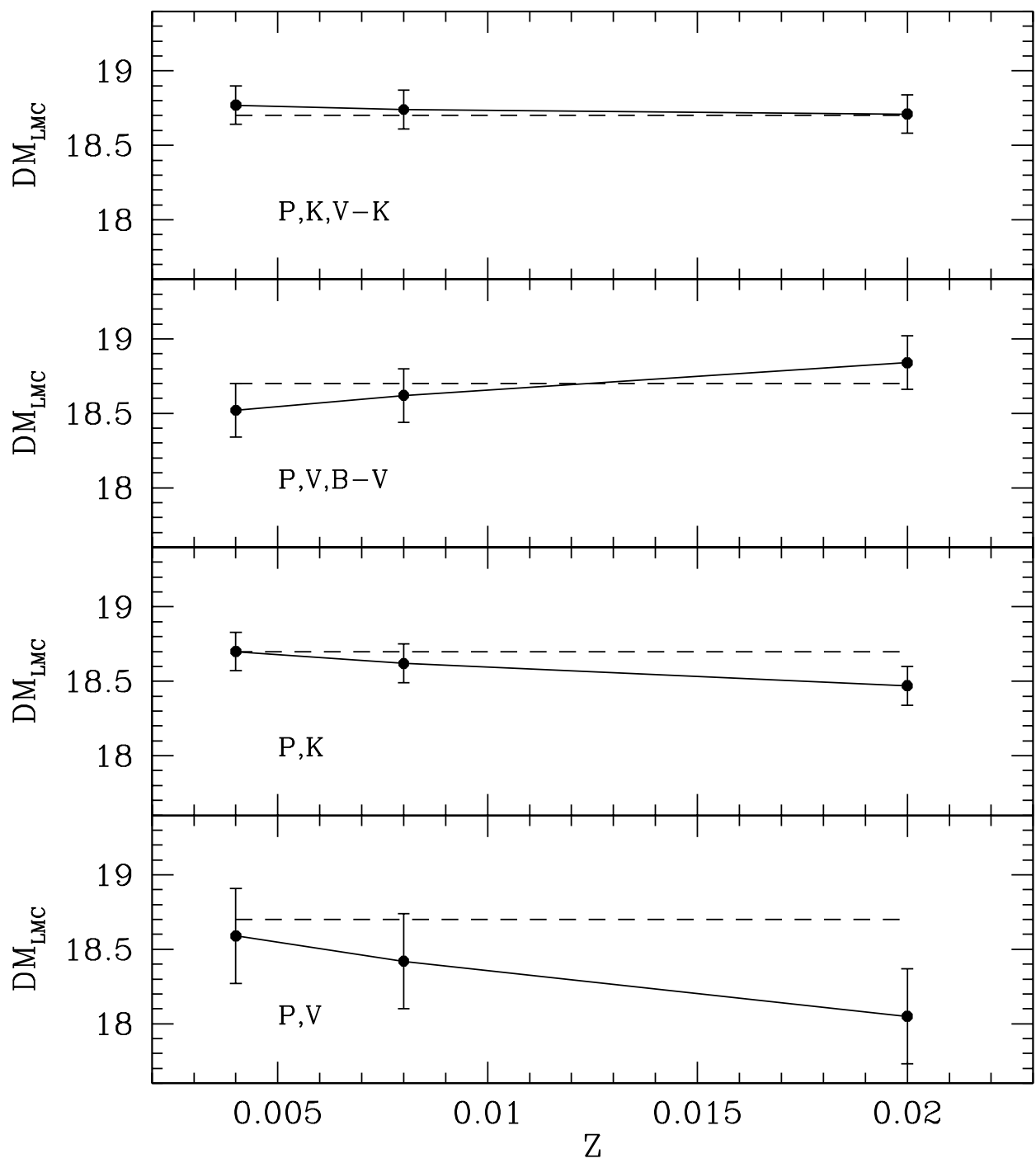
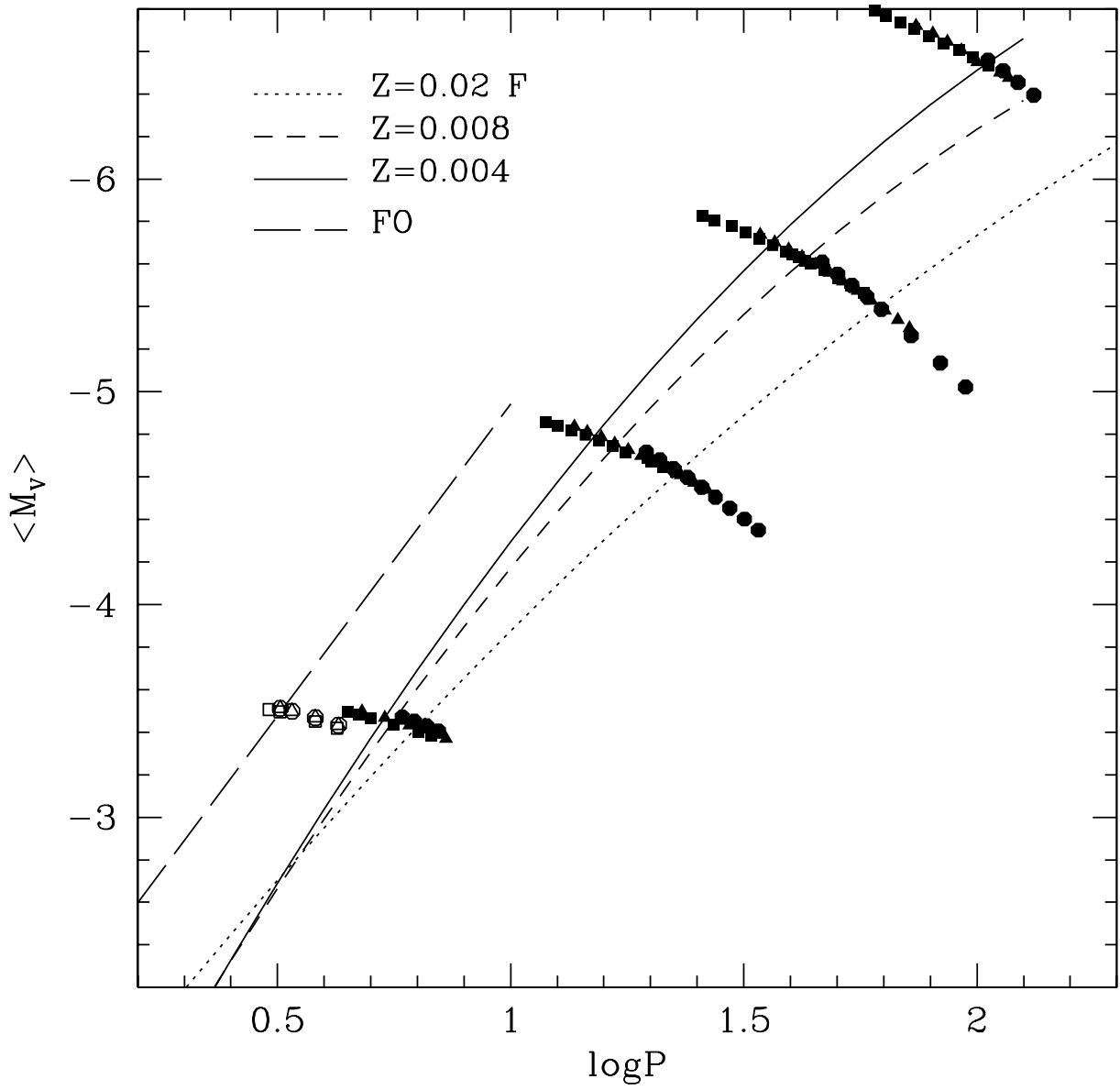
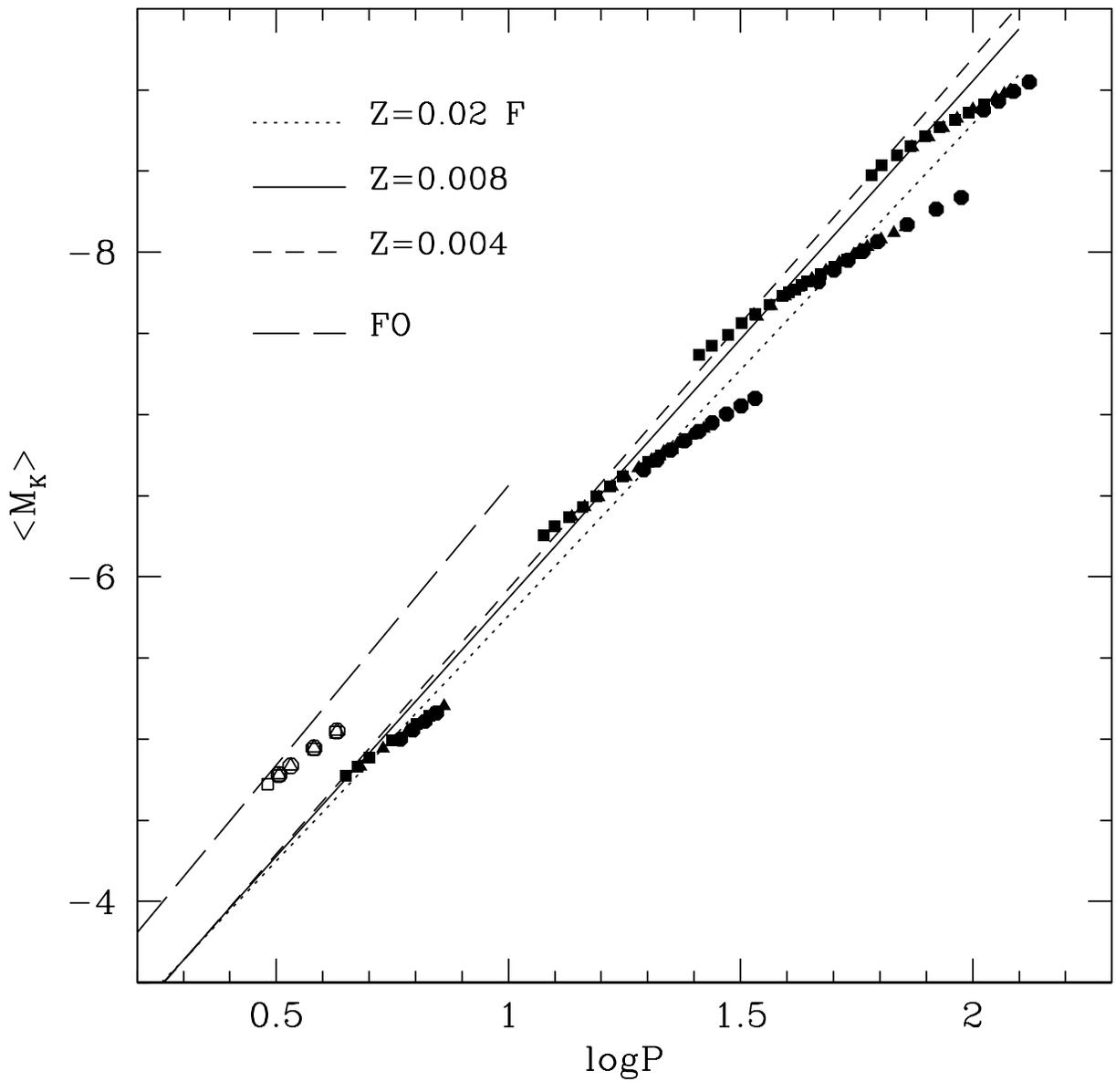
\end{document}